\documentstyle[epsf]{mn} 
\newcommand{\mm}[1]{\mbox{$#1$}} 
\newcommand{\unit}[1]{\ifmmode \:\mbox{\rm #1}\else \mbox{#1}\fi} 
\newcommand{\bv}[1]{\bmath{#1}} 
\renewcommand{\sb}[1]{_{\rm #1}} 
\newcommand{\expec}[1]{\mm{\left\langle #1 \right\rangle}} 
\newcommand{\mone}{\mm{^{-1}}}

\newcommand{\dif}{\mbox{d}} 
\newcommand{\kms}{\unit{km~s\mone}} 
 
\newcommand{\mpc}{\unit{Mpc}} 
 
\newcommand{\hmpc}{\mm{h\mone}\mpc}

\newcommand{\hr}{\mm{^{\rm h}}} 

\newcommand{\lb}[2]{\mm{l = #1\degr}, \mm{b = #2\degr}} 
 
\newcommand{\lbapr}[2]{\mm{l \approx #1\degr}, \mm{b \approx #2\degr}} 
\newcommand{\wrt}{with respect to} 

\newcommand{\dnsig}{\mm{D\sb{n}-\sigma}}

\newcommand{\secref}[1]{Section~\ref{sec:#1}} 
 
\newcommand{\eqref}[1]{equation~(\ref{eq:#1})} 
 
\newcommand{\figref}[1]{Fig.~\ref{fig:#1}} 
\newcommand{\tabref}[1]{Table~\ref{tab:#1}} 
\title {Galaxy clusters in the Perseus--Pisces region --- 
II.\\The peculiar velocity field} 

\author [M.J. Hudson et al.] 
{ 
M.J. Hudson$^{1,2}$\thanks{CITA National Fellow}, 
J.R. Lucey$^1$, 
R.J. Smith$^1$, 
J. Steel$^1$ 
\\ 
$^1$ Department of Physics, University of Durham, South Road, Durham DH1 3LE, 
United Kingdom.\\ 
$^2$ Department of Physics \& Astronomy, University of Victoria, 
P.O. Box 3055, Victoria BC V8W 3PN, Canada. 
} 
\begin{document} 
\maketitle 

\begin{abstract} 
  We have measured the mean peculiar motions of 103 early-type
  galaxies in 7 clusters in the Perseus--Pisces (PP) ridge or PP
  background, and a further 249 such galaxies in 9 calibrating
  clusters from the literature, using the inverse Fundamental Plane
  relation. This relation is found to have a distance error of 20\%
  per galaxy. None of the 6 clusters in the PP ridge has a significant
  motion with respect to the cosmic microwave background (CMB) frame,
  but the PP background cluster J8 shows marginal evidence of
  `backside infall' into the PP supercluster. We find that the full 16
  cluster sample has a mean CMB-frame bulk motion of $420\pm280$
  \kms\, towards \lb{262}{-25}. This result is consistent both with no
  bulk motion in the CMB frame and with the $\sim 350$ \kms\ bulk
  motion found by Courteau, Faber, Dressler and Willick. It is
  inconsistent at the 98\% confidence level with the $\sim 700$ \kms\ 
  bulk flow found by Lauer \& Postman.  The PP ridge clusters are
  found to have a small and statistically insignificant mean radial
  motion with respect to the CMB frame: $-60\pm220$ \kms.  Our error
  analysis fully accounts for the uncertainties in the mean Hubble
  flow, as well as the errors due to the merging of different
  spectroscopic datasets. A comparison between our cluster peculiar
  velocities and the predicted peculiar velocities from the IRAS 1.2
  Jy density field, smoothed on a 500 \kms\ scale, yields $\beta_I
  \equiv \Omega^{0.6}/b_I = 0.95\pm0.48$, consistent with previous
  results. We find agreement between our peculiar motions and
  published Tully-Fisher results for the same clusters. The
  disagreement between the 11 clusters common to our sample and that
  of Lauer \& Postman, based on brightest cluster galaxies (BCGs), is
  statistically significant at the $\ga 99.7$\% confidence level
  indicating that the errors of one or both of these data sets are
  underestimated. When the BCG distances corrected for the X-ray
  luminosity of the host cluster are used, the disagreement is reduced
  to the $\sim$94\% confidence level.

\end{abstract} 


\section{Introduction} 
\label{sec:intro} 

The two dominant concentrations of {\em galaxies\/} in the nearby
Universe are the Great Attractor (hereafter GA) supercluster complex
and the Perseus--Pisces (hereafter PP) supercluster. The GA appears to
be extended along the line of sight at \lbapr{310}{20}, with its
densest region at and behind the Centaurus cluster in the distance
range $\sim 3000$ -- $4500$ \kms\ (Hudson 1993). On the opposite side
of the sky, the main ridge of PP is extended perpendicular to the line
of sight with $4500 \la cz_{\sun} \la 5500$ \kms\ (Giovanelli and
Haynes 1985; Wegner, Haynes and Giovanelli 1993).  The only way to
measure the distribution of {\em mass} on these large scales is by
measuring the peculiar velocities of galaxies and of clusters of
galaxies.

Studies of the peculiar flow field around these dominant
concentrations can be traced from Shaya (1984) and Tammann \& Sandage
(1985) who found that, after having accounted for Virgo infall, nearby
galaxies showed a residual motion with respect to the frame of the
Cosmic Microwave Background (hereafter CMB) in the direction of
Hydra-Centaurus. A tidal shear of the flow field in this direction was
also noted by Lilje, Yahil \& Jones (1986). Strong infall into a Great
Attractor {\em behind\/} the Centaurus cluster at $cz \sim 3000$ \kms\ 
was first claimed by Lynden-Bell et al.\ (1988).  While it is now
evident that there is a strong streaming motion of galaxies in the
direction of Centaurus, it remains unclear whether this motion is
generated locally, e.g.\ by Centaurus or the GA, or whether it is due
to more distant sources. For example, the claim by Mathewson, Ford and
Buchhorn (1992) that `backside infall' is not observed behind the
Great Attractor, argues for coherent streaming motions generated at
very large distances.

In practice, the flow field is likely to be complex, being dominated
by attractors and voids on a range of scales. Nevertheless, to first
order, the mean motion of the PP supercluster allows a test of whether
the flow is due to local or distant sources since, in combination with
peculiar velocity data near the GA, both the bulk motion and the shear
of the peculiar velocity field can be measured. If the large bulk
motion of nearby galaxies is due to local sources, then the peculiar
velocity of PP will be small: $\sim -100 \kms$.  Alternatively, if the
source of the motion is more distant, then PP should participate in a
large-scale coherent bulk flow and have a peculiar velocity of $\sim
-400$ \kms\ with respect to the CMB.

Previous work on the motion of PP has been based mainly on the spiral
galaxy Tully-Fisher (TF) relation, the PP region not being
well-sampled in the elliptical galaxy survey of Faber et al.\ (1989).
Willick (1990, 1991), reported a mean radial motion of $-440\pm50$
\kms\ for PP field spirals with redshifts 3800 $< cz <$ 6000 \kms\ 
based on r-band TF data. The small quoted error on this result
includes only the random error from the TF scatter. The result is
subject also to a systematic calibration error of 2\%, or 100 \kms\ at
PP (Willick 1991). Courteau et al.\ (1993) added further r-band TF
data and found that the PP region participated in a uniform bulk flow
of $360\pm40$ \kms\ towards \lb{294}{0}. Using I-band TF peculiar
velocities of clusters in PP, Han \& Mould (1992, hereafter HM)
claimed a motion of --400 \kms. In contrast, the recent TF field
survey of Giovanelli et al.\ (1996; da Costa et al.\ 1996) indicates
no net motion of PP.

The PP supercluster is at least as overdense in galaxies if not more
so than the GA (Saunders et al.\ 1991; Hudson 1993; Strauss \& Willick
1995). If PP is as massive as implied by its overdensity in galaxies,
then strong infall of galaxies into the PP supercluster is predicted.
The infall signature is difficult to detect unambiguously using field
samples because individual galaxies scatter out of the filament,
creating a spurious infall signature. This inhomogeneous Malmquist
bias (Hudson 1994a; Dekel 1994) is particularly severe in the PP
region due to the sharpness of the density contrast between the
supercluster filament and the foreground and background voids. While
cluster distances are far less severely affected by Malmquist biases,
the clusters typically reside along the ridge line of the PP filament
itself, so that they are not efficient tracers of the infall pattern.

In this paper, we use the Fundamental Plane (hereafter FP) distance
indicator to measure the peculiar velocities of early-type galaxies in
clusters in and behind the PP filament. The use of a cluster sample
minimises the Malmquist bias problems which plague field samples
(Strauss \& Willick 1995). Furthermore, since the cores of clusters
are rich in early-type galaxies, cluster-elliptical samples are less
prone to contamination than spiral samples.

The outline of this paper is as follows. In \secref{data}, we discuss
the cluster and galaxy samples. In \secref{fp}, we derive the FP
relation for our combined data sample. In \secref{dis}, we present the
distances and peculiar velocities of individual clusters, and in
\secref{flow} we model the flow field and measure the bulk motion of
the sample and the mean flow of the clusters in the PP filament.
\secref{sys} discusses potential systematic effects. In \secref{comp}
we compare our results to previous work. Finally, in \secref{summary}
the paper is summarized.

Note that distances -- either the estimated distance $d$ or the true
position $r$ -- are quoted in units of \kms. Radial peculiar
velocities, $u \equiv \bv{v}\cdot\hat{\bv{r}} \equiv cz - d$, are with
respect to the CMB frame unless otherwise noted.

\section{Data} 
\label{sec:data} 

The principal goal of this work is to determine the motion of the PP
supercluster and compare this motion with the predictions of flow
models. In order to derive distances to the PP clusters we must adopt
a distance indicator zero-point. The approach often used is to adopt
one cluster, usually Coma, as a calibrator which is assumed to be at
rest with respect to the CMB. However, individual clusters will have
peculiar velocities with respect to the CMB of order $\sim 300$ \kms\ 
(Grammann et al.\ 1995). For the case of Coma, this would translate to
a zero-point uncertainty of 4\%, and hence to an uncertainty of $\sim
200$ \kms\ in any derived motion of the PP supercluster. In order to
reduce this uncertainty, we use here a set of clusters as calibrators
and simultaneously fit for the distance indicator zero-point and a
flow model. To break the degeneracy between the zero-point and a bulk
motion, we require a sample of clusters with good sky coverage.

In Smith et al. (1997, hereafter PPI), we reported the spectroscopy 
and R-band photometry for early-type galaxies in seven clusters in the 
PP region. This includes the six clusters (7S21, Pisces, HMS0122, 
A262, A347 and Perseus[$\equiv$A426]) in the PP ridge ($4500 \la 
cz_{\sun} \la 5500$ \kms) and the background cluster J8 at $cz_{\sun} 
\sim 10000$ \kms. Our `calibration' clusters are from the published 
studies of Lucey and collaborators (Lucey \& Carter 1988, hereafter 
LC88; Lucey et al. 1997, hereafter LGSC) and J{\o}rgensen, Franx \& 
Kj{\ae}rgaard (1995a, 1995b, 1996, hereafter JFK95a, JFK95b, JFK96). 
We exclude the poor clusters from these studies, i.e. Pavo II, S639, 
S753, Doradus($\equiv$Grm13) and Grm15. The Centaurus cluster is also 
excluded. Centaurus has a complicated internal structure, i.e. the 
Cen30 and Cen45 components, and is located in the GA foreground, so 
that it is not likely to be a good calibrator. This culling restricts 
our calibration sample to nine clusters, viz. A194, A539, 
Hydra($\equiv$A1060), Coma($\equiv$A1656), A2199, A2634, A3381, 
A3574($\equiv$K27) and DC2345-28($\equiv$A4038$\equiv$K44). Details 
of all 16 clusters considered here, including the data sources used, 
are given in \tabref{cluscz}. Note that in our analysis, we solve for 
the distances of all 16 clusters simultaneously and hence the term 
`calibration sample' is somewhat artificial. 
\begin{table*}
\caption{Cluster Sample}
\label{tab:cluscz}
\begin{tabular}{lrrrrrrllll}
Cluster &
\multicolumn{1}{c}{n} &
\multicolumn{1}{c}{$l$} &
\multicolumn{1}{c}{$b$} &
\multicolumn{1}{c}{$cz_{\sun}$} &
\multicolumn{1}{c}{$cz_{\rm CMB}$} &
\multicolumn{1}{c}{$\epsilon_{cz}$} &
Subsample &
Spec. &
Phot. &
Band  \\
~\\
7S21       &  7 & 113.8 & -40.0 &  5860 &  5517 &  189 & PP ridge & PPI & PPI & Kron-Cousins R\\
Pisces     & 25 & 126.8 & -30.3 &  5011 &  4714 &  100 & PP ridge & PPI & PPI & Kron-Cousins R\\
HMS0122    &  9 & 130.2 & -27.0 &  4914 &  4636 &  167 & PP ridge & PPI & PPI & Kron-Cousins R\\
A262       & 10 & 136.6 & -25.1 &  4782 &  4528 &  158 & PP ridge & PPI & PPI & Kron-Cousins R\\
A347       &  8 & 140.7 & -18.1 &  5528 &  5312 &  177 & PP ridge & PPI & PPI & Kron-Cousins R\\
Perseus    & 31 & 150.5 & -13.7 &  5202 &  5040 &  245 & PP ridge & PPI & PPI & Kron-Cousins R\\
J8         & 13 & 150.3 & -34.4 &  9664 &  9425 &  177 & PP backgd. & PPI & PPI & Kron-Cousins R\\
~\\
A2199      & 36 &  62.9 &  43.7 &  8922 &  8947 &  106 & Calib. & LGSC & LGSC & V\\
A2634      & 35 & 103.5 & -33.7 &  9505 &  9158 &  132 & Calib. & LGSC & LGSC & V\\
Coma       & 71 &  57.6 &  88.0 &  6931 &  7200 &  113 & Calib. & LGSC & LGSC & V\\
A194       & 19 & 142.2 & -62.9 &  5428 &  5122 &  115 & Calib. & LC88/JFK95b & JFK95a & Gunn-r\\
A539       & 22 & 195.6 & -17.6 &  8612 &  8615 &  190 & Calib. & JFK95b & JFK95a & Gunn-r\\
A3381      & 14 & 240.3 & -22.7 & 11369 & 11471 &  134 & Calib. & JFK95b & JFK95a & Gunn-r\\
A3574      &  7 & 317.4 &  31.0 &  4604 &  4873 &  189 & Calib. & JFK95b & JFK95a & Gunn-r\\
DC2345-28  & 27 &  25.3 & -75.8 &  8772 &  8473 &  148 & Calib. & LC88 & JFK95a & Gunn-r\\
Hydra      & 18 & 269.6 &  26.5 &  3632 &  3976 &  152 & Calib. & LC88/JFK95b & JFK95a & Gunn-r\\
\end{tabular}
\end{table*}

For the Coma, A2199 and A2634 clusters, we use the spectroscopic and 
V-band photometric data as tabulated in LGSC. The spectroscopic data 
is from several systems, including a few measurements from our PP 
runs, but excludes the erroneous `FLEX' velocity dispersion data of 
Lucey et al. (1991). The data for the remaining 6 rich clusters are 
from LC88 and JFK. For these six clusters we take the spectroscopic 
measurements of LC88 and JFK95b, and the photometry from JFK95a's 
r-band measurements. Our full sample consists of 352 galaxies in 16 
clusters. 

All spectroscopic data have been corrected for aperture effects 
following JFK95b and brought onto a common system using the derived 
offsets and weightings given in PPI. The effective surface 
brightnesses, $\langle\mu\rangle\sb{e}$, are k-corrected, corrected 
for $(1+z)^4$ surface brightness dimming and for Galactic extinction. 
For the latter, we use the Burstein \& Heiles (1982) values of 
$E(B-V)$ multiplied by the factors 2.35, 3.02 and 2.50 to obtain 
extinctions in the R, V and Gunn r bands, respectively. Surface 
brightnesses are transformed from the V and Gunn-r bands to the 
Kron-Cousins R band by subtracting 0.57 magnitudes and 0.33 magnitudes 
respectively (see PPI). Uncertainties in these transformations 
introduce only 0.5\% distance errors and so are neglected. 

 From all sources, we exclude galaxies with morphological type S0/a and 
later or which are classified as morphological rejects (type R) in 
PPI. The following background galaxies were also excluded: A01094 in 
A262 ($cz = 14620$ \kms), ZH56 in A194 ($cz = 8269$ \kms) and RMH30 in 
Hydra ($cz = 10672$ \kms). Finally, the galaxies RB28 in A2199, B03C 
in A347 and R338 in Hydra are outliers (at the $>3.5 \sigma$ level) in 
the FP relation found below. They are excluded from the analysis in 
this paper. 

The recession velocity adopted for each cluster is the mean redshift
of the observed sample of galaxies in the cluster.  In
\tabref{cluscz}, the parameter $\epsilon_{cz} =
\sigma\sb{cl}/\sqrt{n}$ gives the error in the mean sample redshift,
where $\sigma\sb{cl}$ is set to 500 \kms\ for clusters with $n < 10$
members, and is set to the measured cluster dispersion or 500 \kms,
whichever is larger, for clusters with $n \ge 10$. We find that our
mean sample redshifts and published mean redshifts for the same
clusters agree to within $\la \epsilon_{cz}$.

\section{The Fundamental Plane} 
\label{sec:fp} 

The methodology adopted in this paper is to fit cluster FP relations
{\em a priori\/} without the use of redshift information, then to
convert the cluster zero-points into cluster distances and finally to
fit flow models {\em a posteriori\/}. Following Strauss \& Willick
(1995), we refer to this procedure as `Method I'. In this section we
describe the FP fits. The Malmquist bias corrections and cosmological
corrections needed to transform the FP zero-points into cluster
distances are deferred to \secref{dis}.

\subsection{Method of fit} 

The Fundamental Plane can be described by either a `forward' or an
`inverse' fit, depending on whether the slopes are obtained by
regressing on the distance dependent parameter --- the logarithm of
the effective radius, $\log R\sb{e}$ --- or by regressing on the
distance independent parameter, $\log \sigma$. Note that, in the
inevitable presence of selection effects and scatter, the slope of the
inverse relation is not simply the inverse of the forward relation
slope.

Galaxies in cluster samples are typically selected according to their
morphological properties and either surface brightness, magnitude,
diameter or some combination of these. The inverse relations have the
advantage that they are unbiased by selection on the dependent
variables.  For the FP, this means that any selection in $R\sb{e}$,
$\langle\mu\rangle\sb{e}$ or any function of these quantities (e.g.\ 
total magnitude) will not bias the inverse FP fit.

For the PP clusters, galaxies were selected from photographic plates,
and are limited by APM or GSC photographic magnitudes. From the
candidate lists, spectra were obtained typically from brightest galaxy
on down. There are, however, cases were the photographic magnitude was
re-estimated by eye (e.g.\ due to contamination by a nearby star or
galaxy) or where fainter galaxies were preferentially observed because
they could be placed on the same spectrograph slit as a brighter
galaxy. Therefore, the selection criteria are likely to be somewhat
fuzzy and are difficult to quantify precisely. As a result it is
difficult to apply the forward-relation bias-correction scheme
described by Willick (1994). Because the inverse distance indicator
relations are insensitive to photometric selection, we shall consider
only the inverse FP distance indicator in this paper. Of course the
inverse relation is sensitive to {\em explicit\/} selection on
velocity dispersion, but such selection is not present in our sample
because we have not thrown out any galaxies {\em a posteriori\/} based
on their velocity dispersions. There is one exception to this rule:
the Coma sample of LGSC explicitly excludes all galaxies with $\sigma
< 100$ \kms. For this cluster we make a bias correction for the
$\sigma$ cut which is exactly analogous to the bias correction
prescription of Willick (1994) for the calibration of the forward
Tully-Fisher relation from a magnitude-limited sample of cluster
galaxies. In practice, this correction has only a very small effect
due to the large range of $\sigma$ in the Coma sample.

The inverse FP relation used here is:
\begin{equation} \label{eq:dieq4} 
\log \sigma = \frac{1}{\alpha} \log R\sb{e} - \frac{\beta}{\alpha} 
\langle\mu\rangle\sb{e} - \frac{1}{\alpha} \gamma\sb{cl} 
\end{equation} 

We minimise the $\log \sigma$ residuals over the galaxies in all
clusters simultaneously, assuming the same slopes ($\alpha$, $\beta$)% 
\footnote{Our definition of $\beta$ differs by a factor --2.5 from that 
of JFK96. Here $\beta$ is the coefficient of $\langle\mu\rangle_{e}$ whereas 
for JFK96 it is the coefficient of $\log\expec{I}_e$} for all 
clusters but allowing individual zero-points ($\gamma_{\rm cl}$) for each 
cluster to vary independently. In order to determine the errors on 
the zero-points, the slope of the FP relation are held 
fixed. Note that our slopes, derived from an inverse fit, will not be 
the same as those obtained from the ``orthogonal fit'' of JFK96. 

\subsection{Results of FP fits} 

\begin{table}
\caption{Fit parameters and scatters for the distance indicator relations}
\label{tab:fitparms}
\begin{tabular}{lcc}
& FP & \dnsig\ \\
$\alpha$ &	$ 1.383\pm 0.040$ & $ 1.419\pm 0.044$ \\
$\beta$ &	$ 0.326\pm 0.011$ & -- \\
 $\Delta\sb{\sigma}$ &	0.062 & 0.065 \\
$\Delta\sb{inv}$ &	0.198 & 0.213 \\
\end{tabular}
\end{table}

\begin{figure*}\epsfxsize=180mm 
\epsfbox{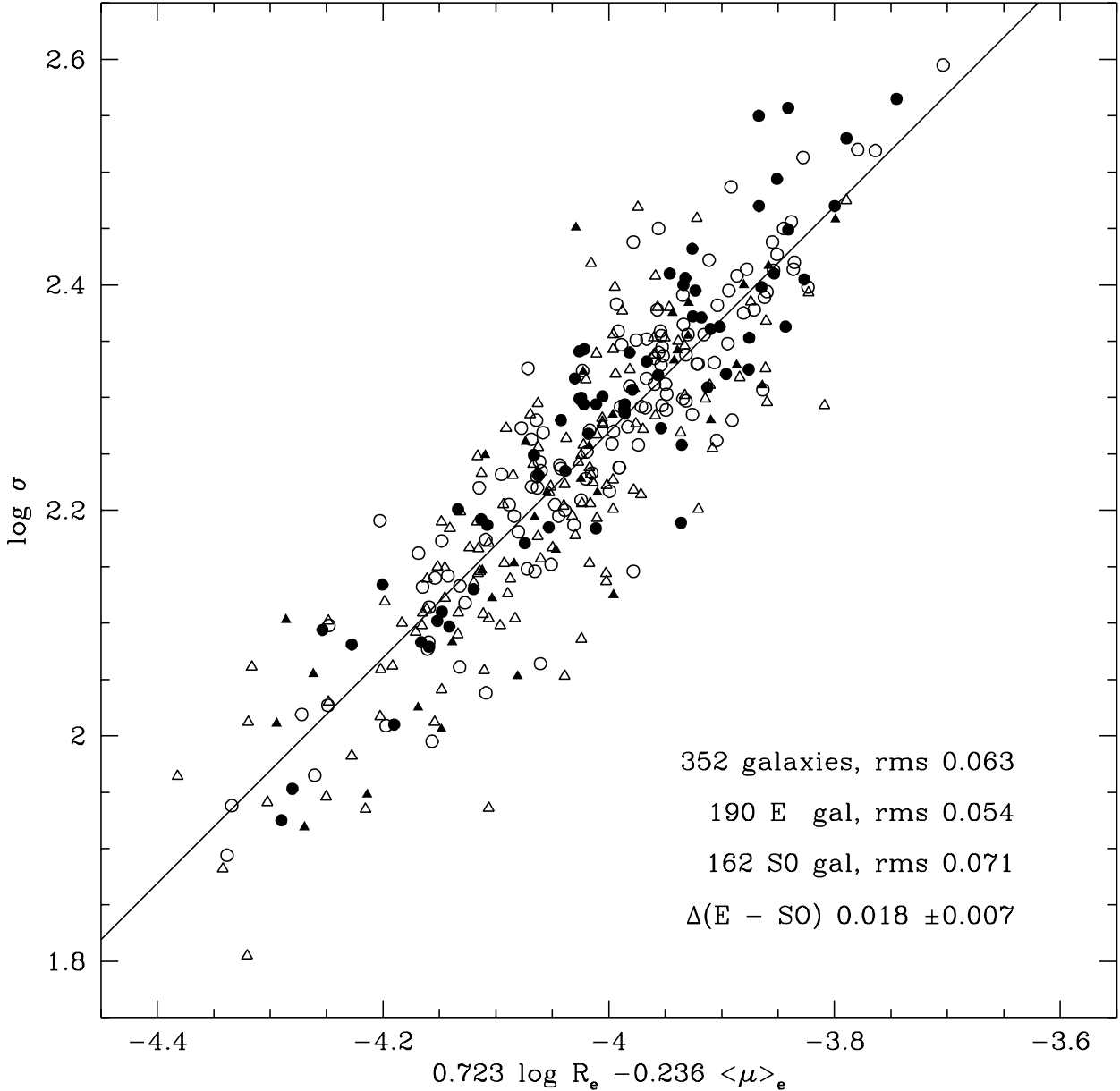} 
\caption{ 
The FP relation for all galaxies. All clusters are shifted by their 
relative mean CMB redshifts, so that they are placed at the distance 
of Coma. Early type galaxies (E, E/S0, D or cD) are indicated by 
circles, later types are indicated by triangles. The 103 galaxies in 
the 7 PP clusters are represented by filled symbols. The remaining 
249 galaxies in the 9 calibrating clusters are shown by the open 
symbols. The rms scatter in $\log \sigma$, $\Delta\sb{\sigma}$, for 
the full sample, and the E and S0 subsamples is given in the lower 
right hand corner, with the offset between E and S0 galaxies in the 
inverse FP relation. See \secref{morph} for further discussion.} 
\label{fig:fpall} 
\end{figure*}

The FP relation for the 16 cluster sample is shown in \figref{fpall}.
Here, the mean CMB redshifts of the clusters have been used to shift
all galaxies to the distance of the Coma cluster.

\tabref{fitparms} gives the slopes, scatter in velocity dispersion,
$\Delta_{\sigma}$, and the fractional inverse distance error,
$\Delta\sb{inv} \equiv \alpha\,\ln(10)\,\Delta_{\sigma} $, for the FP
relation. Note that the distance error per galaxy is larger than that
found from the orthogonal fits of JFK96, through the influence of the
larger slope, $\alpha$. For comparison, we give also the slope and
scatter of the inverse \dnsig\ relation, for the same data. This
distance indicator has slightly larger scatter than the FP, and is not
used in the analysis to follow.

\begin{figure*}\epsfxsize=180mm 
\epsfbox{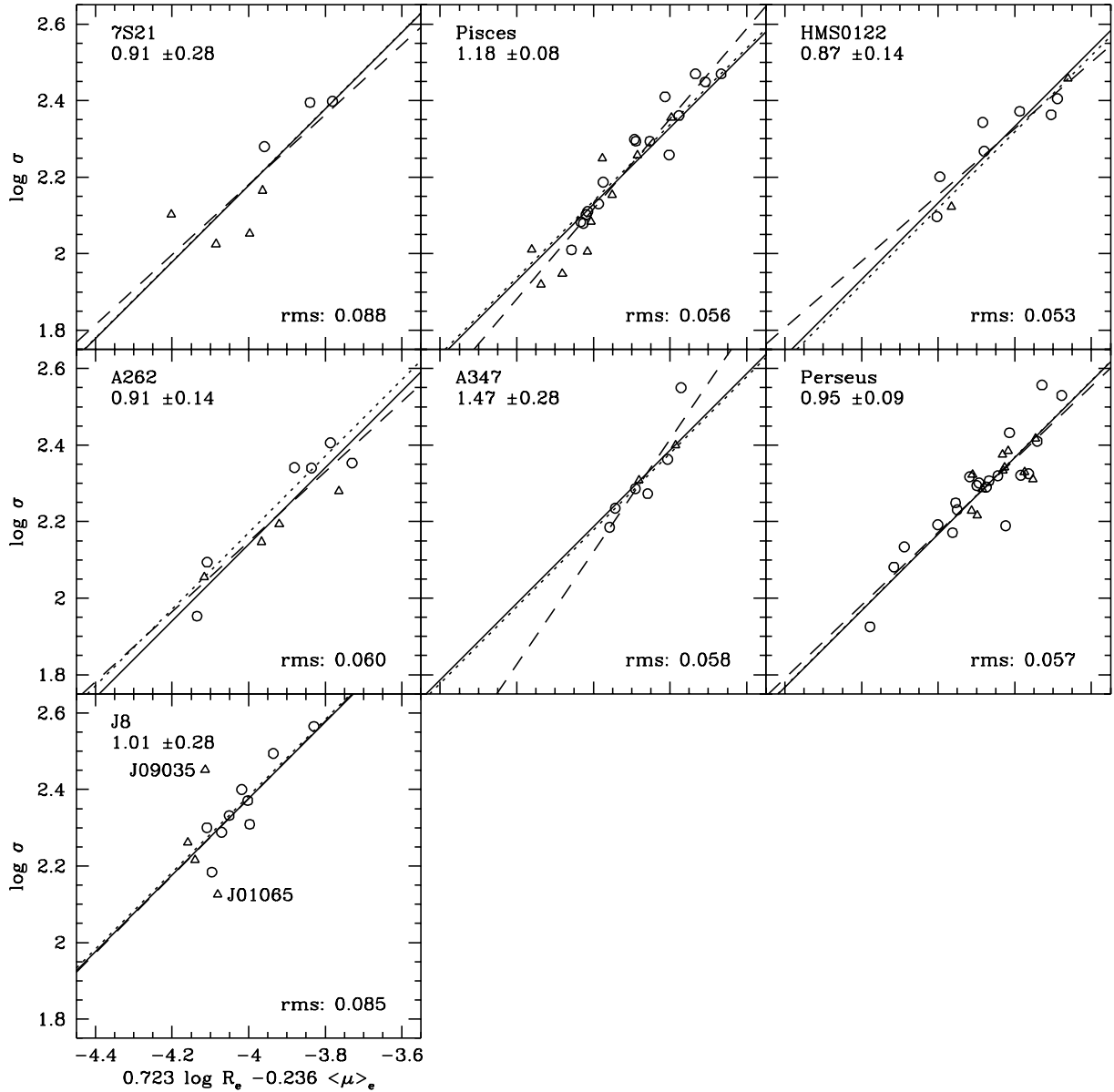} 
\caption{ 
  FP data and fits for the PP clusters 7S21, Pisces, HMS0122, A262,
  A347, Perseus and J8. Early type galaxies (E, E/S0, D or cD) are
  indicated by circles, later types are indicated by triangles. The
  solid line shows the global inverse FP, found by minimising $\log
  \sigma$ residuals simultaneously over the whole cluster sample with
  the same slope but varying zero-points for each cluster. Each
  cluster's measured scatter in $\log \sigma$ around this global fit
  is given in the lower right-hand corner. Galaxies which deviate from
  the global fit by more than 2.5 times the global scatter are
  labelled. The dotted line shows the median of the residuals with the
  slope fixed from the whole cluster sample. The dashed line shows the
  best slope and zero-point fit to the individual cluster. The
  individual cluster slope relative to the global slope is given in
  the upper left-hand corner.
\label{fig:fp1}} 
\end{figure*}

\begin{figure*}\epsfxsize=180mm 
\epsfbox{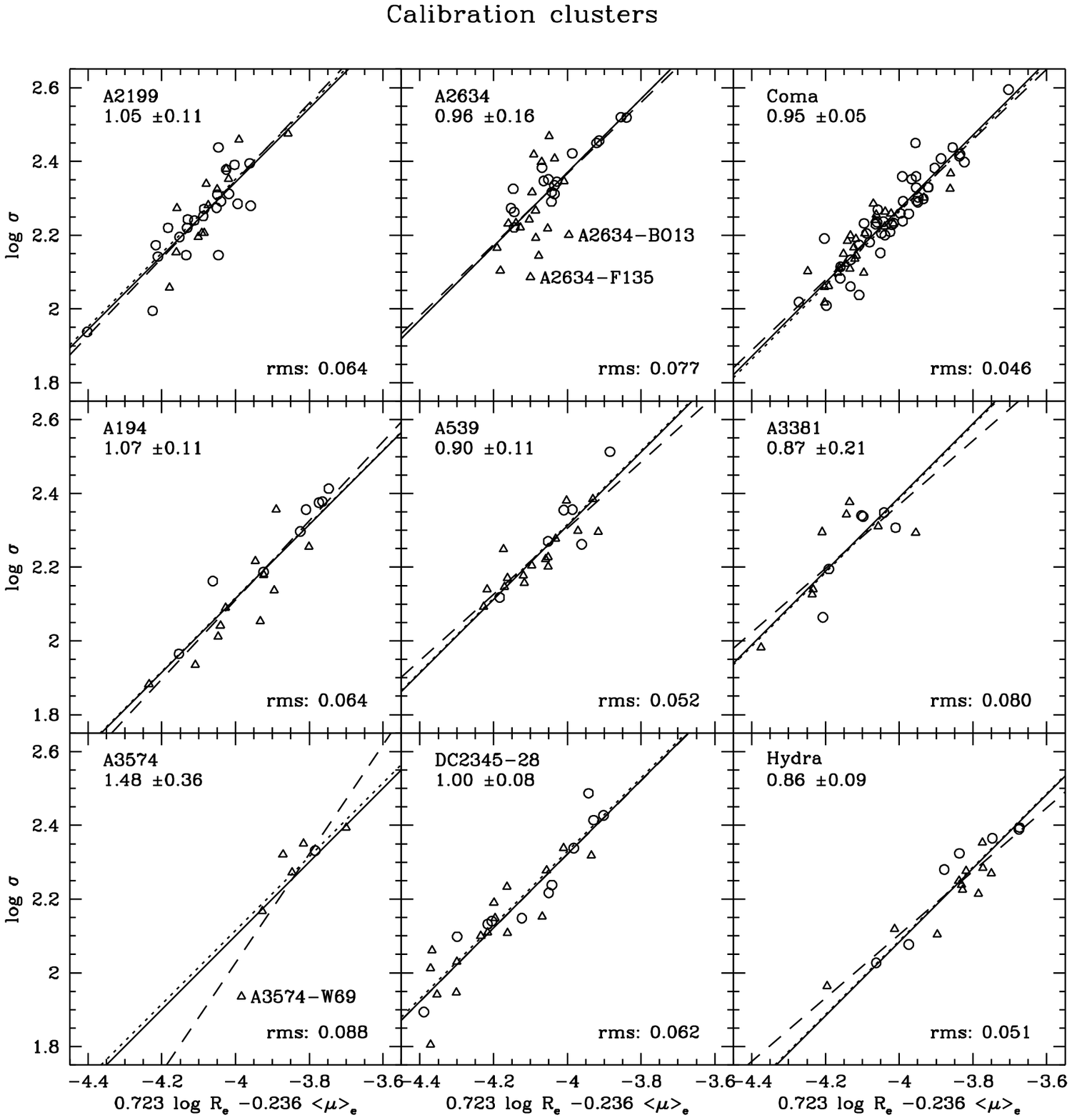} 
\caption{ 
  FP data and fits for the clusters A2199, A2634, Coma, A194, A539,
  A3381, A3574, DC2345-28 and Hydra.  Symbols and curves are as in
  \figref{fp1}.
\label{fig:fp2} 
} 
\end{figure*}

Figures~\ref{fig:fp1}--\ref{fig:fp2} show the data for each cluster in
separate panels. The solid line shows the distance indicator relation
using the slope derived from all galaxies. The dotted line shows the
median fit to the residuals. The difference between the mean and
median zero-point is always $\la 1\sigma$, with the exception of A262
for which they differ at the $1.6 \sigma$ level. The scatter around
the mean (solid) line is indicated in the lower right-hand corner of
each panel. The individual cluster scatters are all consistent with
the global scatter, with the exception of Coma, the scatter for which
is marginally smaller: 0.046 versus the global scatter 0.062.  This
difference is only marginally significant (at the $\sim 95$\%
confidence level). Note that since Coma is traditionally the standard
cluster for FP/\dnsig\ studies, its galaxies have been observed many
times. Thus their mean velocity dispersions will have smaller random
errors than galaxies in less frequently observed clusters. The dashed
line in each panel shows the individual fit with free slope to the
cluster in question. The slopes (relative to the slope of best fit)
are given in the top left hand corner. None of these slopes are
significantly different from unity, except for Pisces, for which the
difference is marginal ($\sim 2\sigma$).

\section{Distances and Peculiar velocities} 
\label{sec:dis} 

\subsection{Bias corrections} 
We argued in \secref{fp} that the inverse fits were insensitive to
selection on $R\sb{e}$, $\langle\mu\rangle\sb{e}$ or functions
thereof. As discussed above, we perform in this paper a Method I
peculiar velocity analysis. Whenever a Method I procedure is used, the
distances will be affected by Malmquist bias, which depends on the
density field of the peculiar velocity tracer population, before
selection.  For a Method I analysis using forward distance indicator
relations, this would be the only source of bias.  In contrast, a
Method I analysis using inverse relations is affected, in general,
both by Malmquist bias and by `object selection bias'.  For samples of
{\em field\/} galaxies, the object selection bias is strong,
particularly at the distance where the selection function peaks and
beyond. For samples consisting of rich clusters, as is the case here,
the probability that an {\em entire cluster\/} will drop out is a very
weak function of distance and hence the object selection bias is
negligible, leaving only the Malmquist bias. For further discussion of
these issues the reader is referred to Strauss \& Willick (1995).

One of the advantages of a cluster sample is that the corrections for
Malmquist bias are smaller than for field galaxies. The fractional
distance error for each cluster is $\Delta\sb{inv}/\sqrt{n}$, where
$n$ is the number of observed galaxies in the cluster. The homogeneous
Malmquist correction, which assumes that the underlying density field
of clusters is uniform, is a multiplicative factor
$\exp(-3.5\Delta\sb{inv}^2/n)$. For our sample, the largest
homogeneous Malmquist correction is 2\% (or $\sim 100$ \kms\ at the
distance of the PP ridge) for the cluster 7S21 which has 7 observed
members. The correction for the Perseus and Pisces clusters, which
dominate the flow by virtue of their small random errors, is only
$\sim 0.5\%$ or $\sim 25$ \kms. The correction due to the fact that
the underlying cluster density field is inhomogeneous (Hudson 1994a;
Dekel 1994; Strauss \& Willick 1995) is expected to be smaller than
the homogeneous Malmquist correction, which itself is quite small, as
demonstrated. We therefore neglect the inhomogeneous correction in
this paper.

\subsection{Distance calibration} 

In order to obtain cluster distances, we shall initially fix the
distance and peculiar velocity of Coma to be 7200 \kms\ and zero,
respectively, and calculate cluster distances relative to Coma. In the
flow model fits to follow, we will allow this zero-point to be
rescaled by a free parameter. We shall see below that this free
parameter is close to unity, thus justifying our calibration {\em a
  posteriori\/}.

The difference between the zero-point of a cluster and that of Coma is
the logarithm of relative apparent angular diameters of galaxies (at
the same $\sigma$). We iteratively solve the angular diameter distance
equation with $q_0=0.5$ in order to convert the zero-point offsets to
distances, i.e. the $cz$ that the cluster would have if it were
following the Hubble flow. For $q_0=0$, the distance of clusters in
the PP ridge changes by only 10 \kms.

\subsection{Results} 

\begin{table}
\caption{FP Cluster Distances and Peculiar Velocities}
\label{tab:clusfpdis}
\begin{tabular}{lrr@{$\pm$}rr@{$\pm$}rr}
Cluster &
\multicolumn{1}{c}{$cz_{\rm CMB}$} &
\multicolumn{2}{c}{$d_{\rm MC}$} &
\multicolumn{2}{c}{$u_{\rm CMB}$} &
\multicolumn{1}{c}{$u_{\em IRAS}$} \\
&
\multicolumn{1}{c}{\small \kms} &
\multicolumn{2}{c}{\small \kms} &
\multicolumn{2}{c}{\small \kms} &
\multicolumn{1}{c}{\small \kms} \\
7S21       &  5517 &  5448 &  409 &    69 &  450 &	 -201 \\
Pisces     &  4714 &  4583 &  182 &   131 &  208 &	  108 \\
HMS0122    &  4636 &  4680 &  310 &   -44 &  352 &	  159 \\
A262       &  4528 &  4787 &  300 &  -259 &  340 &	  226 \\
A347       &  5312 &  5589 &  392 &  -277 &  430 &	 -210 \\
Perseus    &  5040 &  5176 &  185 &  -136 &  307 &	  180 \\
J8         &  9425 & 10457 &  576 & -1032 &  602 &	 -706 \\
A2199      &  8947 &  9289 &  307 &  -342 &  325 &	  -29 \\
A2634      &  9158 & 10118 &  339 &  -960 &  364 &	 -127 \\
Coma       &  7200 &  7200 &  170 &     0 &  204 &	  -64 \\
A194       &  5122 &  4379 &  199 &   743 &  230 &	  295 \\
A539       &  8615 &  8381 &  355 &   234 &  403 &	  644 \\
A3381      & 11471 & 10893 &  578 &   578 &  593 &	  -80 \\
A3574      &  4873 &  4218 &  316 &   655 &  369 &	 -126 \\
DC2345-28  &  8473 &  8647 &  330 &  -174 &  362 &	 -129 \\
Hydra      &  3976 &  3946 &  185 &    30 &  239 &	  147 \\
\end{tabular}
\end{table}

The Malmquist-corrected FP distances, $d\sb{MC}$, and radial peculiar
velocities, $u\sb{CMB}$, are given in Table \ref{tab:clusfpdis}. The
distance error, $\epsilon_d$, is due to the error in fitting the
zero-point. The peculiar velocity error, $\epsilon_u$, is the distance
error added in quadrature with the cluster mean redshift error,
$\epsilon_{cz}$.

\begin{figure*}\epsfxsize=180mm 
\epsfbox{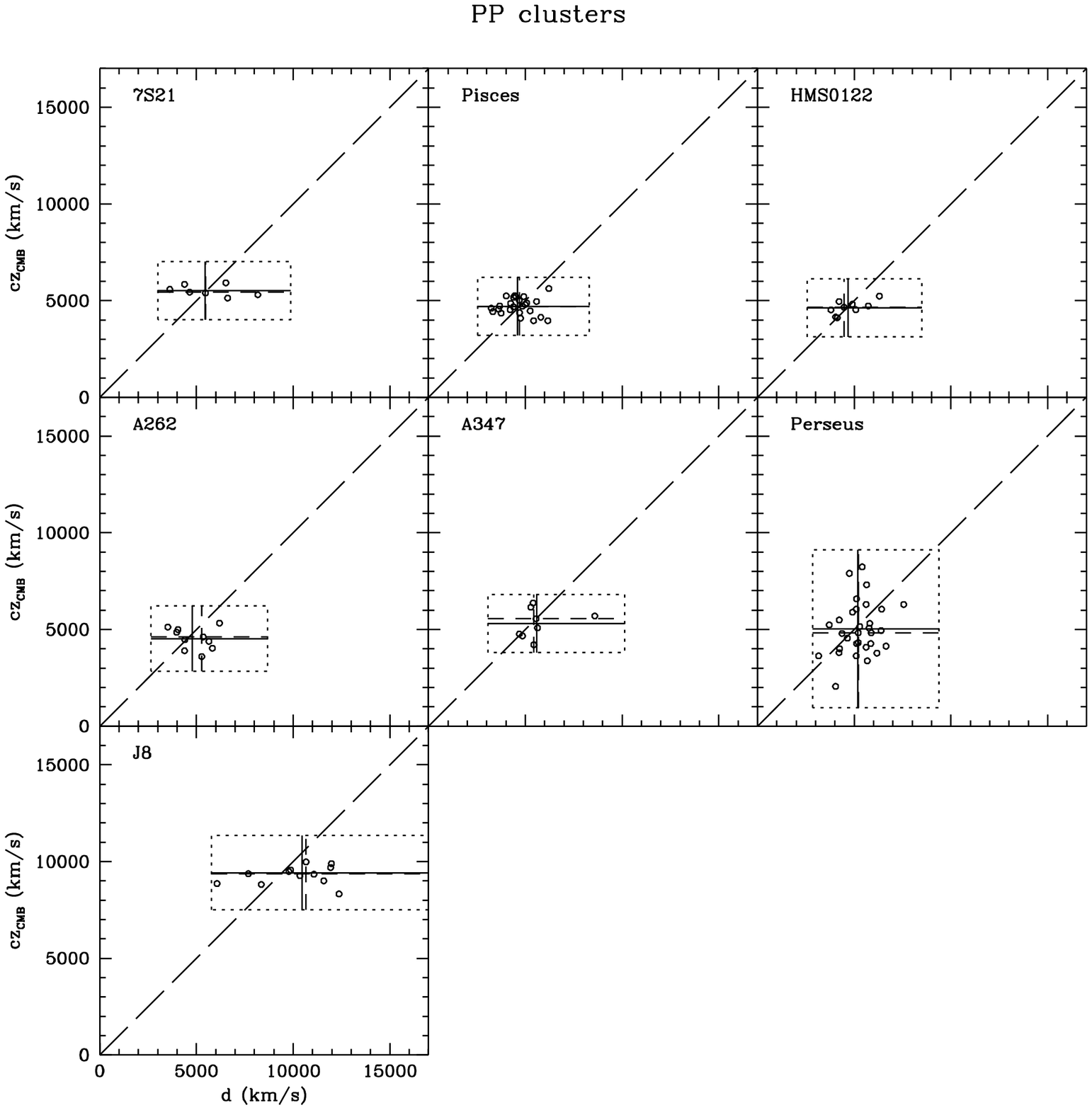} 
\caption{ 
  Distances and redshifts of individual galaxies in 7S21, Pisces,
  HMS0122, A262, A347, Perseus and J8. Horizontal and vertical solid
  (dotted) lines give the mean (3$\sigma$ range) of redshift and
  log(distance), respectively. The short dashed lines give the
  respective medians. The Hubble flow is shown the diagonal
  long-dashed line. The offset of the cross from the Hubble line gives
  the cluster peculiar velocity.}
\label{fig:czd1} 
\end{figure*}

\begin{figure*}\epsfxsize=180mm 
\epsfbox{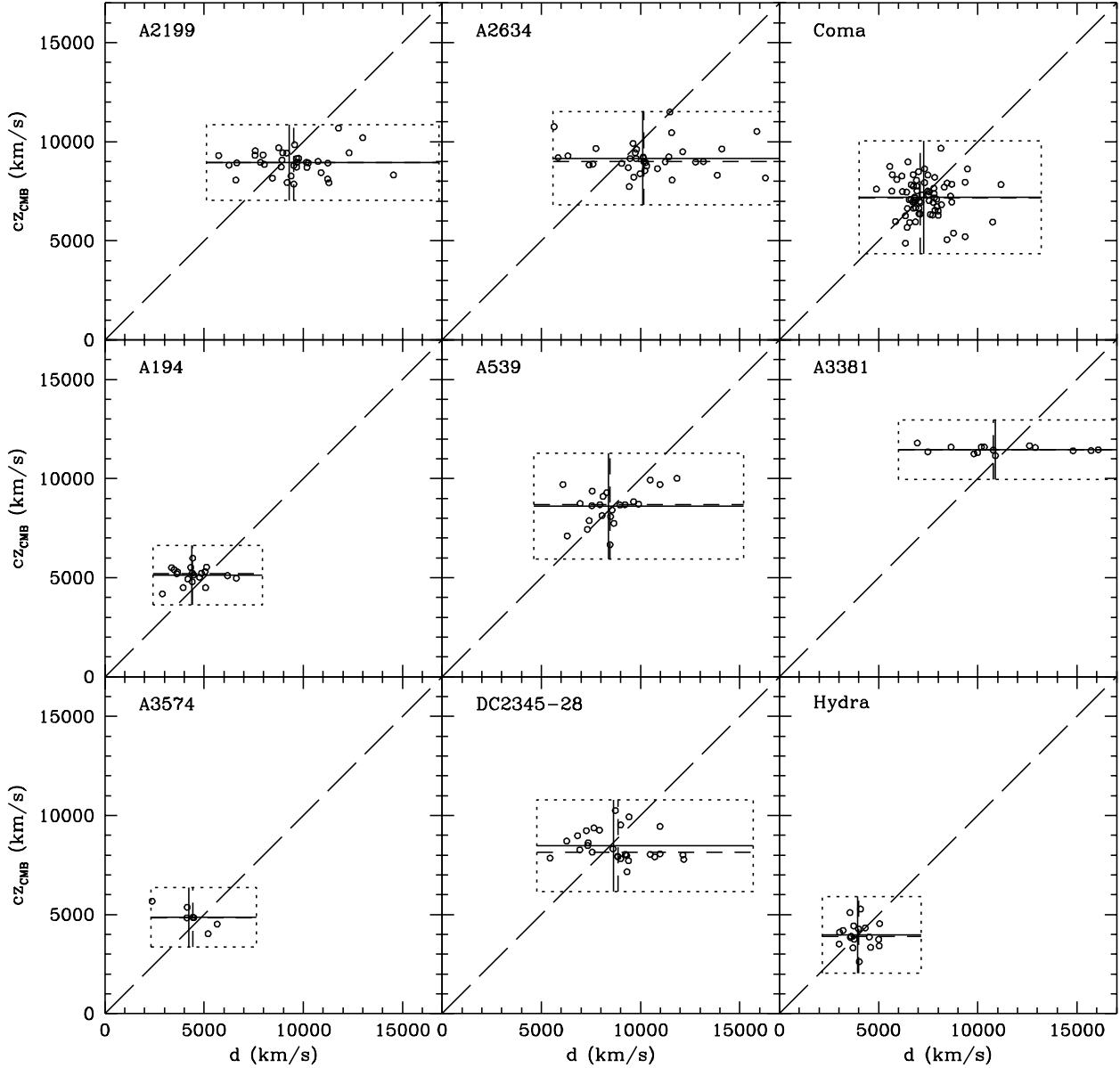} 
\caption{ 
  As \figref{czd1} for the clusters A2199, A2634, Coma, A194, A539,
  A3381, A3574, DC2345-28 and Hydra.}
\label{fig:czd2} 
\end{figure*}

Figures \ref{fig:czd1} and \ref{fig:czd2} show the CMB redshifts and
the FP-inferred distances of individual galaxies in each cluster. Note
that there is no tendency for galaxies to lie along the Hubble line,
in contrast with some spiral-cluster samples (Willick et al.\ 1995).
By observing early-type galaxies, we have efficiently selected
galaxies in the virialized cluster cores.

\begin{figure}\epsfxsize=85mm 
\epsfbox{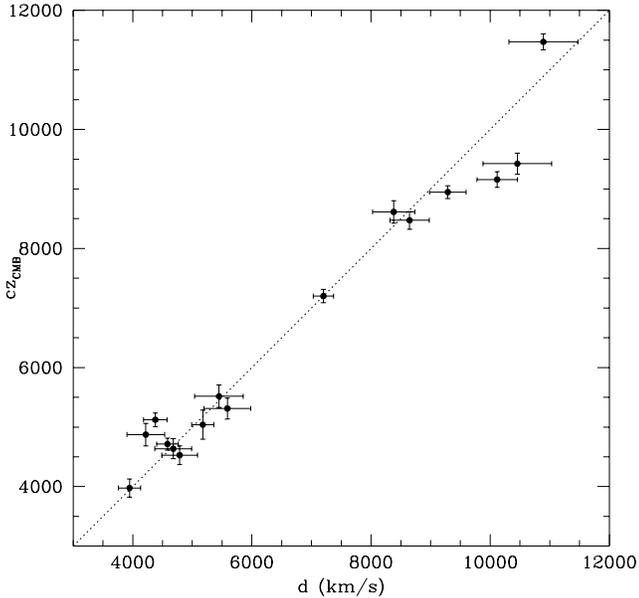} 
\caption{ 
  Hubble diagram for all 16 clusters.}
\label{fig:hubble} 
\end{figure}

The Hubble diagram for all 16 clusters in shown in \figref{hubble}.
Note that most clusters have a peculiar velocity in the CMB frame
which is not different from zero by $>2\sigma$. The exceptions are
A2634 and A194.

\begin{figure*}\epsfxsize=180mm 
\epsfbox{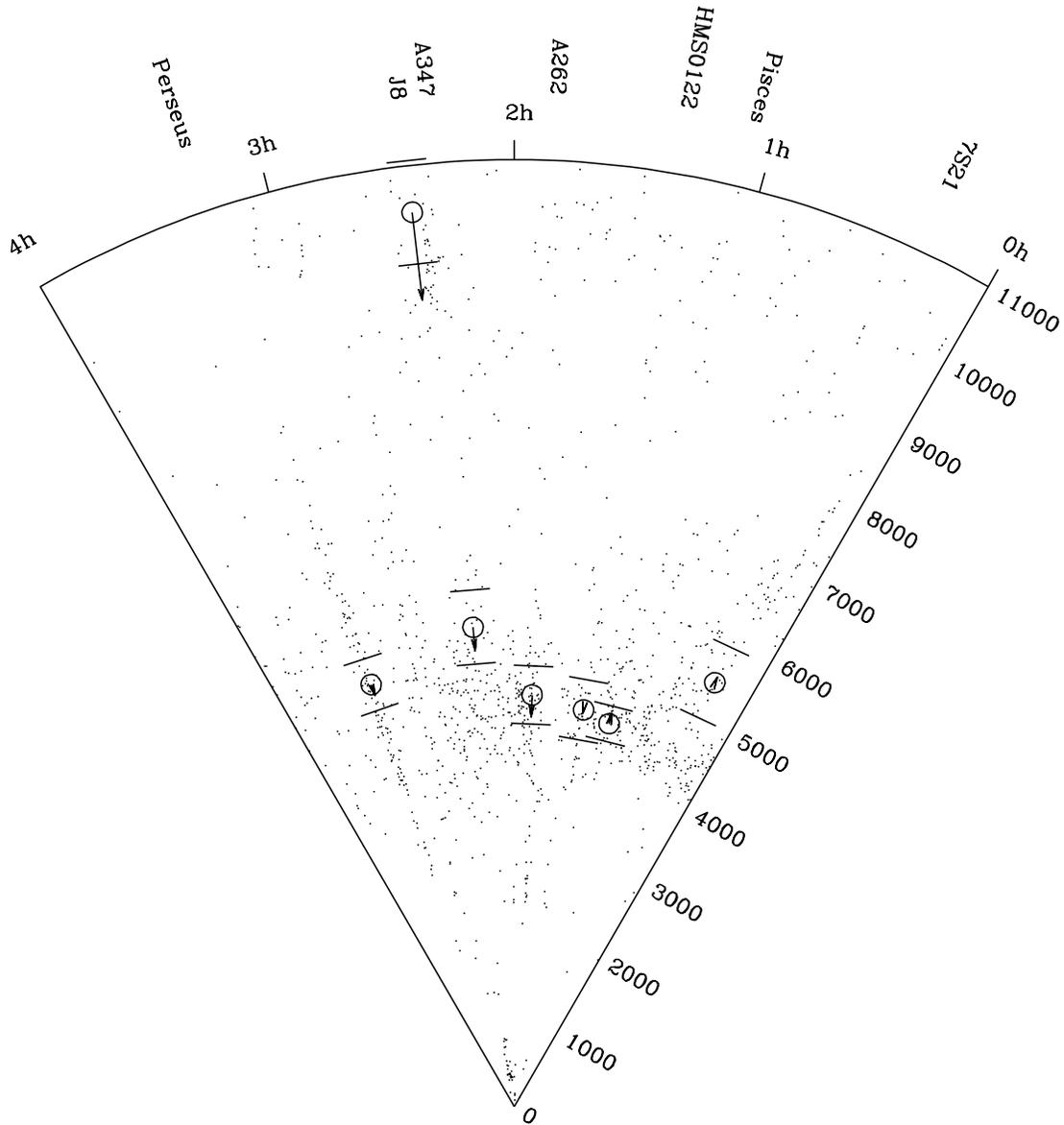} 
\caption{ 
  Flow field for PP clusters with the open circle indicating the FP
  distance of the cluster and the straight lines showing the 1$\sigma$
  distance error. The tips of the arrow show the mean $cz$ in the CMB
  frame. Also displayed are the CMB redshift space positions of
  galaxies taken from the CfA ZCAT compilation. Note that the
  `cluster' of points just west of J8 is a thin wall which runs
  north-south and is seen here in projection.}
\label{fig:bigrapie} 
\end{figure*} 

\begin{figure*}\epsfxsize=180mm 
\epsfbox{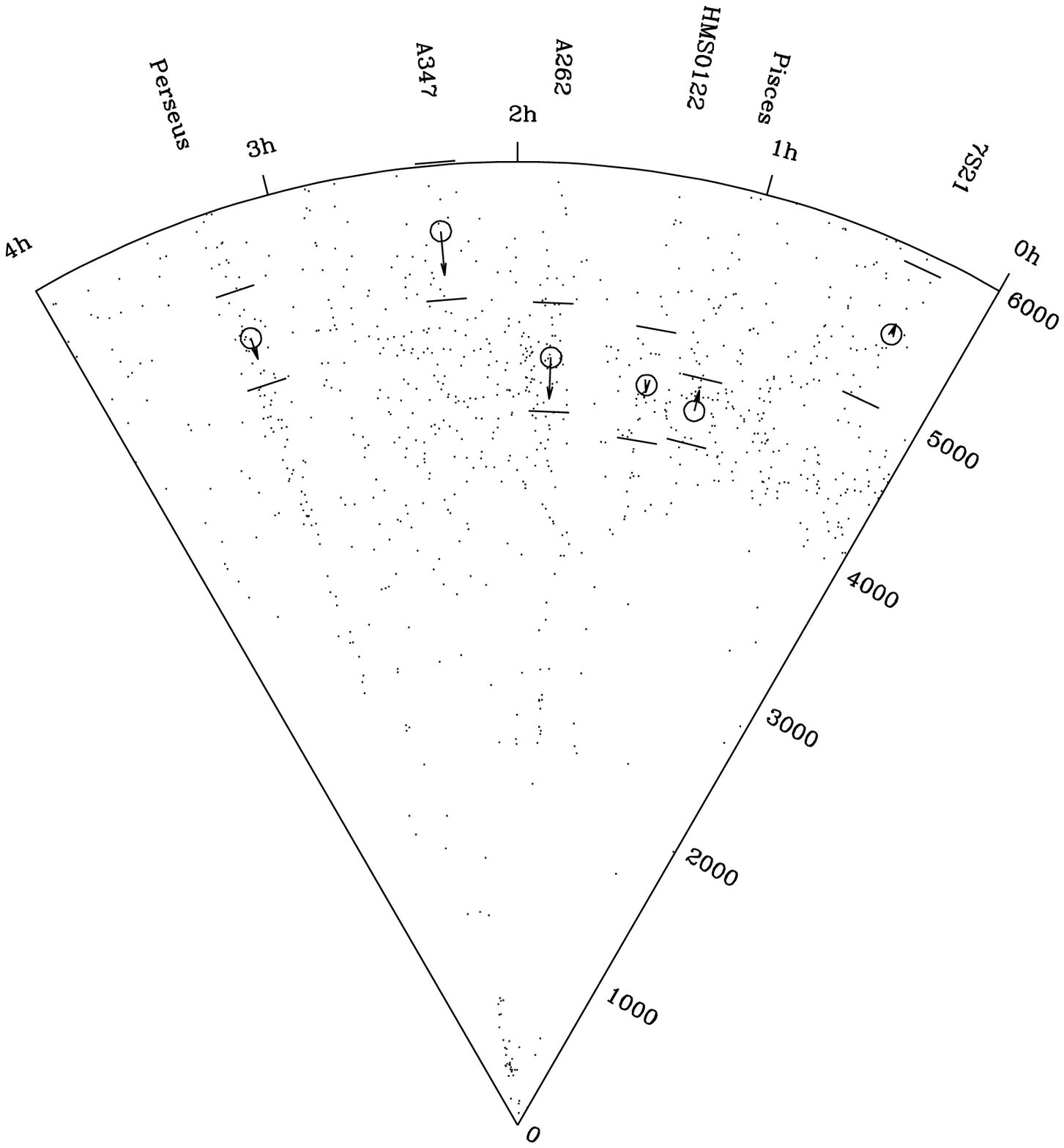} 
\caption{ 
  As in \figref{bigrapie}, but only out to a distance of 6000 \kms, to
  show more detail in the PP ridge.}
\label{fig:smallrapie} 
\end{figure*} 

Figures \ref{fig:bigrapie} and \ref{fig:smallrapie} show the flow
field of the PP clusters studied here. The large circles show the FP
distance to each cluster, while the tip of each arrow lies at its mean
redshift in the CMB frame. The length of the vector thus gives the
peculiar velocity. The horizontal lines bracketing each cluster
position indicate the distance errors. The small points are plotted at
the (redshift-space) positions of galaxies from the CfA ZCAT
compilation (Huchra et al.\ 1993), and are shown to illustrate the PP
structure as traced by galaxies. It is clear from these figures that
the mean CMB-frame motion of the PP ridge is small, and that the
motion of PP ridge clusters with respect to the mean motion of the
ridge is quiet. Note also that the cluster J8 shows some evidence of
backside infall into the PP supercluster.

\section{The flow field} 
\label{sec:flow} 

\subsection{Flow models} 
In order to model the peculiar velocities, we will consider flow
models composed of a combination of pure Hubble flow, a uniform bulk
flow and/or the peculiar velocity predicted from the IRAS 1.2Jy
redshift survey (Fisher et al.\ 1995a).

In order to allow for a possible `thermal' component in the peculiar
velocity of Coma, our adopted calibrating cluster, we allow the
zero-point of the distance scale to float by including a monopole (or
Hubble) term
\begin{equation} 
\bv{v}\sb{H}(\bv{r}) = (1 + \Delta\sb{H}) \bv{r} \, . 
\end{equation} 

The bulk flow is modelled as a uniform dipole motion, independent of 
distance: $\bv{V}$. 

If light traces mass, then the most reasonable flow models are those
in which the galaxy density field is used to predict peculiar
velocities.  There is usually one free parameter for these models,
$\beta \equiv f(\Omega)/b \simeq \Omega^{0.6}/b$. This allows us to
fit a degenerate combination of $\Omega$, which affects the peculiar
velocity field for a given {\em mass} fluctuation field, and the
biasing parameter $b \equiv \delta\sb{gal}/\delta\sb{mass}$ which
relates the observed density fluctuation field of galaxies to that of
the mass. Note that different tracers of the galaxy density field
(e.g.\ IRAS galaxies versus optically-selected galaxies) will have
different bias parameters and hence different values of $\beta$.

In this paper, we use predicted peculiar velocities derived from the
IRAS 1.2~Jy redshift survey density field, which was kindly provided
to us by M. Strauss. The IRAS density field extends to 12000 \kms,
which allows us to make predictions for all of the clusters in our sample% 
\footnote{All-sky density fields of optical galaxies (e.g. Hudson 
1993; Santiago et al.\ 1995 ) are not sufficiently deep for our 
purposes.}. 
Rich clusters of galaxies have collapsed from scales $\sim 8 \hmpc$, 
so we expect that their peculiar velocities are accurately described 
by the predictions of linear theory: 
\begin{equation} 
\bv{v}_I(\bv{r}) = \beta_I \int \delta_I(\bv{r'}) \frac{\bv{r'} - 
\bv{r}} {|\bv{r'} - \bv{r}|^3} \dif^3\bv{r'} \, , 
\label{eq:predpv} 
\end{equation}% 
where $\delta_I$ is the IRAS density fluctuation field, smoothed with
a 500 \kms\ radius Gaussian filter and processed with a Weiner filter
in order to reduce the effects of shot noise.  The smoothing scale is
approximately equivalent to an 8\hmpc\ top-hat, and so is a good match
to the scale from which the clusters collapsed. Note that linear
theory has been used to transform iteratively the IRAS galaxy
positions from redshift space to real space. The $\beta_I$ adopted for
this iterative scheme is $1.0$, which is not significantly different
from the value obtained from fits to our peculiar velocity data below.
Note that the effect of the $\beta$ used in the redshift-to-real space
iteration procedure on the derived $\beta$ is typically very small
(Hudson 1994b; Hudson et al.\ 1995).

Errors in the IRAS predicted peculiar velocities can be estimated from
the results of Fisher et al.\ (1995b). Those authors generated mock
IRAS 1.2 Jy surveys from N-body simulations. Although they used a
different redshift-to-real space reconstruction scheme (based on a
spherical harmonic decomposition), in common with the case here they
also used the Weiner filter to suppress shot noise. We therefore
expect their IRAS predictions to have similar error properties to
those used here.  Fisher et al.\ estimated that the errors in the
IRAS-predicted radial peculiar velocities of individual galaxies were
$\sim 200$ \kms\ within 6000 \kms, growing to approximately 250 \kms
at 11000 \kms. We expect that the errors for predicted peculiar
velocities of clusters should be somewhat smaller than these values.

In general, the predicted peculiar velocity is then 
\begin{equation} 
\bv{v}\sb{p}(\bv{r}) = \bv{v}\sb{H}(\bv{r}) + \bv{V} + 
\bv{v}_I(\bv{r}) \, . 
\end{equation} 
Note that for those cases in which we use the IRAS predictions as part
of the flow model, $\bv{V}$ should be interpreted not as the mean bulk
flow of the sample, but as a {\em residual\/} bulk flow due to mass
fluctuations beyond the limit of the IRAS density field.

Given the model predictions at the estimated distances of each
cluster, we minimise, with respect to the free parameters
$\Delta\sb{H}$, $\bv{V}$ and $\beta_I$, the statistic
\begin{equation} 
\chi^2 = \sum_i^N \frac{(\bv{v}\sb{p}(\bv{d}_i)\cdot\hat{\bv{d}_i} - 
u_i)^2} {\epsilon_{u,i}^2+\epsilon\sb{th}^2} \, , 
\label{eq:chi} 
\end{equation} 
where $\epsilon\sb{th}$ allows for an additional `thermal' peculiar 
velocity of the cluster with respect to the flow model. We set 
$\epsilon\sb{th} = 200$ \kms, so that the reduced $\chi^2$ values are 
approximately unity. Note that we neglect errors in the IRAS predicted 
peculiar velocities, but these are expected to be 
$\la 200$ \kms. The error in the measured radial peculiar velocity is 
typically twice this value; we therefore expect that the derived 
$\beta_I$ will not be significantly biased if errors in the IRAS 
predictions are neglected. 

The formal random errors are obtained from the covariance matrix in
the usual way. This procedure, however, neglects an important source
of error, namely the systematic effects introduced by errors in
matching the different velocity dispersion systems. In PPI, we
determined the corrections which were applied to the raw $\log \sigma$
measurements to bring them onto a common system, and described the
bootstrap resampling method which was used to determine the
uncertainty in these corrections. In order to quantify the effect that
these uncertainties have on our results, we perform the following
procedure.  For each of 200 bootstrap samples, we obtain a realization
of $\log \sigma$ offsets. These are then used to generate new merged
spectroscopic data sets for the 352 galaxies in this paper. The new
data are then passed through the flow analysis above. In this way, we
determine the covariance matrix which is due solely to system matching
errors.  This covariance matrix is then added in quadrature to the
random error covariance matrix to obtain the total error covariance
matrix.  The bulk motion errors quoted in the following subsection
include these systematic contributions.

\subsection{Results} 
\label{sec:results} 

\begin{table*}
\caption{Bulk flow fits to the sample of all 16 clusters}
\label{tab:clusufit}
\begin{tabular}{lccrrrr}
Row &
$\beta$ &
$\Delta_H$ &
\multicolumn{1}{c}{$|V|$} &
\multicolumn{1}{c}{$l$} &
\multicolumn{1}{c}{$b$} &
\multicolumn{1}{c}{$\chi^2$/d.o.f.} \\
1 & 0.00  &  $-0.002\pm0.017$  &  $  424\pm  283$  &  262.6  &  -25.3  &  15.51/12 \\	
2 & 0.00  &  0.000  &  $  430\pm  278$  &  264.7  &  -25.5  &  15.52/13 \\	
3 & $0.95\pm0.48$  &  $0.001\pm0.017$  &  $  389\pm  294$  &  313.2  &  -26.4  &  11.63/11 \\	
4 & $0.95\pm0.48$  &  0.000  &  $  383\pm  271$  &  312.7  &  -26.5  &  11.63/12 \\	
\end{tabular}
\end{table*}

\begin{table*}
\caption{Bulk flow eigenvalues and eigenvectors}
\label{tab:eigen}
\begin{tabular}{lrrrrrrrrrrr}
Row &
\multicolumn{1}{c}{$|V_1|$} &
\multicolumn{1}{c}{$l_1$} &
\multicolumn{1}{c}{$b_1$} &
\multicolumn{1}{c}{$|V_2|$} &
\multicolumn{1}{c}{$l_2$} &
\multicolumn{1}{c}{$b_2$} &
\multicolumn{1}{c}{$|V_3|$} &
\multicolumn{1}{c}{$l_3$} &
\multicolumn{1}{c}{$b_3$} &
\multicolumn{1}{c}{$\chi^2$} &
\multicolumn{1}{c}{$\cal P$} \\
1  & $   21\pm  134$ & 310.7 &  50.7 & $  423\pm  283$ & 264.6 & -29.5 & $   26\pm  312$ & 188.7 &  23.4 & 2.27 & 0.52 \\
2  & $   42\pm  127$ & 312.7 &  46.9 & $  391\pm  275$ & 288.3 & -40.5 & $  174\pm  299$ & 209.1 &  12.4 & 2.47 & 0.48 \\
3  & $   83\pm  137$ & 305.9 &  51.0 & $  341\pm  286$ & 281.4 & -36.4 & $  167\pm  347$ &  20.5 & -12.1 & 2.01 & 0.57 \\
4  & $  106\pm  130$ & 308.2 &  47.2 & $  357\pm  275$ & 294.8 & -42.0 & $   89\pm  334$ &  30.9 &  -6.7 & 2.42 & 0.49 \\
\end{tabular}
\end{table*}

We fit the bulk flow and IRAS-predicted flow models to two samples of
clusters: the full sample of 16 clusters, and the 6 PP ridge clusters
alone. \tabref{clusufit} lists the values of $\beta$, $\Delta\sb{H}$
and the amplitude and direction of bulk flow $\bv{V}$ for the full
cluster sample. For the first two parameters, entries without a quoted
error indicate that the value was held fixed in the fit. The $\chi^2$
of the fits and the number of degrees of freedom are given in the last
column.

When the covariance matrix of the bulk flow components is
diagonalized, we find that one direction has much smaller errors than
the others. Hereafter this direction will be referred to as the DME
for ``direction of minimum error''. This effect occurs because the
fits are dominated by the PP clusters, which are concentrated in a
small patch on the sky. The rows of \tabref{eigen} give, for each
solution in the corresponding row in \tabref{clusufit}, the directions
of the three eigenvectors of the covariance matrix and the projections
of the bulk motion along the these directions. The last columns give
the significance of bulk flow compared with zero: $\chi^2$ (to be
compared with three degrees of freedom) and the corresponding
probability that the there is no bulk motion in the CMB frame. The
first eigenvector, denoted by the subscript ``1'' is the DME. For the
full cluster sample, the DME is \lb{351}{51}, which is within 20\degr\ 
of the of the bulk motion found by Lauer \& Postman (1994, hereafter
LP) and is close to the directions of the GA (Faber \& Burstein 1988)
and the Shapley Concentration (Raychaudhury 1989). Thus, despite the
large errors in the other eigendirections, this sample can be used to
test the predictions of a number of interesting flow models.

The first row of \tabref{clusufit} indicates that the mean bulk motion
of the entire sample with respect to the CMB frame is $420\pm280$
\kms\ towards \lb{262.6}{-25.3}, which is not significantly different
from zero.

The fits with a free Hubble term, $\Delta\sb{H}$, (Rows 1 and 3 in
\tabref{clusufit}) indicate that the best fitting $\Delta\sb{H}$ is
very small and consistent with zero. Thus, by chance, it would appear
that Coma has a small peculiar velocity with respect to the best
fitting flow model. We therefore fix $\Delta\sb{H} \equiv 0$ for the
fits to the PP ridge sample.

For the PP ridge sample, the axis of the DME runs between
\lbapr{314}{26} and \lbapr{134}{-26}, i.e. along the mean radial
direction of the PP clusters, near A262.  The total peculiar velocity
error of 220 \kms\ in this direction can be broken down as 161 \kms\ 
(random error), 125 \kms\ (system matching error) and 85 \kms\ (the
error arising from uncertainty in the distance scale zero-point from
column 2, row 1 of \tabref{clusufit}). Thus the system-matching error
accounts for 18\% of the total error. The errors on the components of
the mean motion in the two other transverse directions are very large
($\ga 700$ \kms\ and $\ga 7000$ \kms), so the motion of the PP ridge
is only well-determined in the radial direction.  The mean radial
peculiar motion of the PP ridge clusters is $-60\pm220$ \kms.  The PP
ridge is thus consistent with having no radial peculiar motion in the
CMB frame.

\begin{figure}\epsfxsize=85mm 
\epsfbox{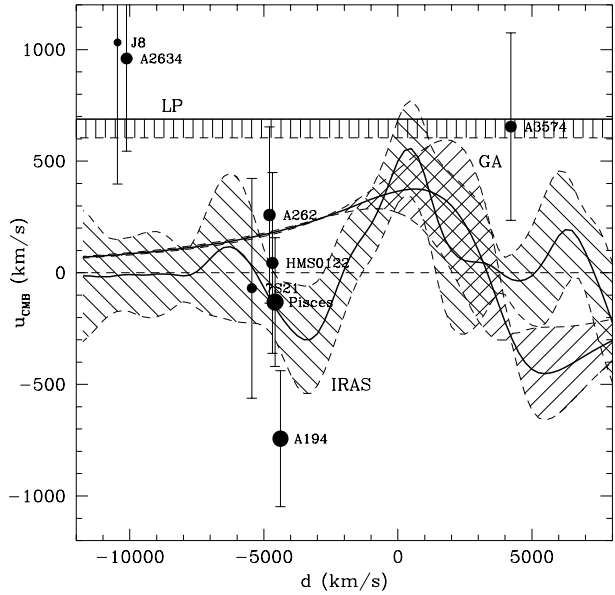} 
\caption{ FP radial peculiar velocities for clusters within 30\degr\ 
  of the direction of minimum error (DME), \lb{311}{51}, as a function
  of their distance from the LG. The distances and peculiar velocity
  of clusters on the Perseus-Pisces side of the sky, i.e.\ opposite to
  the DME, have been multiplied by $-1$. Symbol area is proportional
  the effective weight of the cluster in the fit. The hatched regions
  shows the $\pm1\sigma$ range of peculiar velocities within the
  30\degr\ semi-angle cone predicted by various flow models; the heavy
  solid line shows the mean predicted peculiar velocity within the
  cone. The flow models are labelled as followed: LP -- the bulk
  motion of Lauer \& Postman (1994); GA -- Faber \& Burstein (1988).
  }
\label{fig:cone} 
\end{figure} 

\figref{cone} shows the peculiar velocities projected along the DME as
a function of their distance from the LG (with the PP direction taken
to be negative and the GA direction positive). Only clusters within
30\degr\ of the DME are shown. The predictions of the GA, LP and IRAS
flow models are shown schematically by the hatched regions, which
indicate the $\pm 1\sigma$ range of predicted peculiar velocities
within the cone. Note that the GA and IRAS predictions overlap in the
region where the GA model was first defined, $-2000 < d < 4000$ \kms.
The IRAS predictions are more complex elsewhere, in particular infall
around PP is expected. Note that the PP ridge clusters do not lie in
the foreground or background infall regions, but right along the
centre of mass of the supercluster where $\bv{v}_I \sim 0$. Finally,
note that this diagram is intended to show the qualitative behaviour
of the data and flow models near the DME. It should not be used to
compare the observed and predicted peculiar velocities of individual
clusters since the predictions are only the mean within the cone and
not the value at the position of the cluster itself.

\begin{figure}\epsfxsize=85mm 
\epsfbox{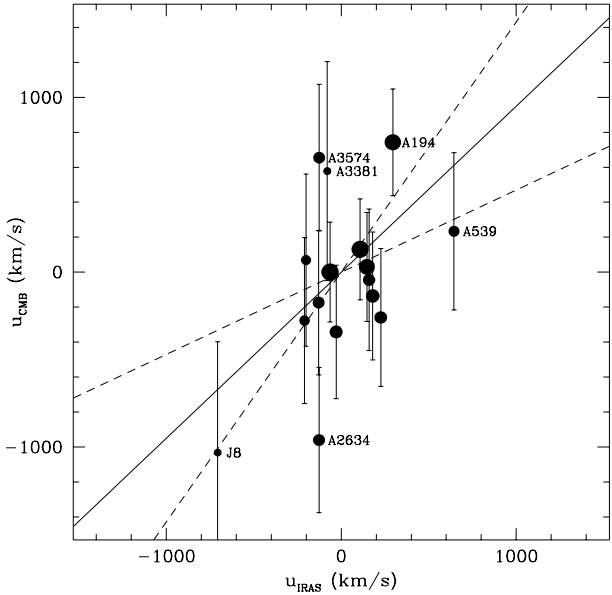} 
\caption{Observed and IRAS-predicted peculiar velocities of individual 
  clusters. The predictions are scaled to $\beta_I = 1$ so the slope
  is a measure of $\beta_I$. Clusters with observed or predicted
  peculiar velocities larger than 500 \kms\ are indicated. The solid
  line shows the slope of best fit, and dashed lines give the
  $1\sigma$ error range. }
\label{fig:iras} 
\end{figure} 

We now consider in more detail the predictions of the IRAS density
field.  \figref{iras} plots the observed peculiar velocities of
individual clusters against the IRAS predictions at their estimated
positions, with $\beta_I = 1$. The slope of the line of best fit thus
gives the value of $\beta_I$: the result is $0.95\pm0.48$. This is a
marginally significant detection of $\beta_I$, but the errors are too
large to place useful constraints on its value. In principle, however
the comparison of cluster peculiar velocity fields with the
predictions from a deep all-sky redshift survey should lead to a
determination of $\beta_I$ which is more reliable than most other
methods. The advantage of our approach is that inhomogeneous Malmquist
bias, smoothing effects and the effects of non-linear bias and gravity
are all small for cluster peculiar velocities.

For the fits with $\beta_I \ne 0$, the quoted bulk motion is the {\em
  residual\/} CMB-frame motion due to sources not modelled by the IRAS
density field, i.e.\ beyond 12000 \kms\ or due to shot noise errors
within that distance (see Willick et al.\ 1997b for a discussion of
these errors). It is interesting to note that the prediction for the
error-weighted mean motion of the PP ridge clusters is $60 \beta_I$
\kms\ {\em away\/} from the LG. This has a direction opposite to the
prediction of the GA model. Presumably the effects of other
structures, such as the foreground void, are important for the
dynamics of the PP ridge. The residual motions of both the full sample
and the PP ridge sample are small and consistent with zero. Finally,
we note that the IRAS-predicted peculiar velocity for J8 is $-706 $
\kms\ in good agreement with the observed value $-1032\pm602$ \kms.
However, for A2634, the IRAS prediction is only $-127$ \kms, which
appears to be marginally inconsistent with the observed $-960\pm364$
\kms.  Note, however, that the shot-noise error in the IRAS prediction
is $\sim 250$ \kms\ and the systematic uncertainty in the peculiar
velocity of A2634 is $\sim 230$ \kms.

% GA vs bulk flow 
We have considered fits to the GA model of Faber \& Burstein (1988).
Our sample is not ideal, however, for constraining the GA model, since
it contains few clusters in the regions where the GA model predicts
strong infall. In particular the Cen30 and Cen45 clusters have been
excluded from our sample. In performing fits of this type, we find
poor agreement with the model. The disagreement does not arise from
the small infall motion of the PP ridge, which is in rough agreement
with the GA model, but is instead partly due to A3574 which has a
peculiar velocity of $655\pm369$ \kms\ near the centre of the GA,
where its peculiar velocity should be $\sim 0$ \kms. This result
should be taken with some caution since the peculiar velocity of A3574
may be biased by a low $\sigma$ galaxy which is an outlier in the FP
relation (see \secref{lowsig} below).

% Shapley 
We have also considered models involving infall to the Shapley
Concentration (Raychaudhury 1989). We find that the results are
statistically indistinguishable from a pure bulk motion in the same
direction, because we have no clusters close enough to the Shapley
Concentration for its tidal shear to be important.

% Thermal 
Finally, we can use our results to constrain the thermal component of
the clusters' peculiar velocity with respect to a given flow model.
We do this by varying $\epsilon\sb{th}$ and requiring that the
resulting $\chi^2$ (\eqref{chi}) lie in an acceptable range. For the
full cluster sample, we find that $\epsilon\sb{th} < 550$ \kms\ at the
95\% confidence level, with the most likely value being $\sim 250$
\kms.  This agrees well with the typical values seen in N-body
simulations (Gramann et al.\ 1995). The flow field of the 6 clusters
in the PP ridge seems to be particularly cold: the 95\% upper limit on
their rms peculiar velocity is $\sim 150$ \kms.

\section{Tests of potential systematic effects} 
\label{sec:sys} 

Systematic errors are often a major concern for distance measurements.
In the analysis above, we have considered the uncertainty introduced
by errors in matching the different velocity dispersion datasets. In
this section we explore other effects that might lead to biases in our
peculiar velocity results.

\subsection{Effect of low-dispersion galaxies} 
\label{sec:lowsig} 

Galaxies with velocity dispersions below 100 \kms\ can have large
systematic and random errors in the measured velocity dispersion
(JFK95b). Our PP target sample contains only 5 galaxies with velocity
dispersions smaller than 100 \kms, and the full sample of 16 clusters
contains only 16 such galaxies. We have tested the effect these
galaxies might have on our solutions by explicitly excluding them, and
correcting for the $\sigma$ cut in a manner exactly analogous to the
bias correction prescription of Willick (1994) for the calibration of
the forward Tully-Fisher relation from a magnitude-limited sample of
cluster galaxies. We find that none of our results (the slopes of the
FP relation, the peculiar velocities of individual clusters or the
global flow model fits) change by an amount larger than the random
error, with the exception of the peculiar velocity of A3574, which
drops from 655 \kms\ to 66 \kms\ when the galaxy W69 is excluded.

\subsection{Effects of galaxy morphology} 
\label{sec:morph} 

We have assigned galaxies to morphological categories using their
appearance on our CCD frames, and their radial profiles derived from
the aperture photometry. The classifications for PP galaxies is given
in PPI. Note that the categories are broad: E ; S0; Q (unclassified or
unclassifiable; there are only two galaxies with this classification).

To assess the validity of the equal treatment of E and S0 galaxies in
our analysis, we can compare the relative offset of elliptical
galaxies (E, E/S0, cD or D) and S0s (S0, S0/E) with respect to the
mean inverse FP relation. Elliptical galaxies have slightly higher
$\log \sigma$ at the same values of $R\sb{e} - 0.326
\langle\mu\rangle\sb{e}$ than S0 types by $0.018\pm0.007$. Using a
sample which has a large number of galaxies in common with that
considered here, JKF96 found a difference of $0.006\pm0.011$ in the
opposite sense. Our sample includes about twice as many galaxies as
that considered by JFK and we have adopted an optimal correction
scheme for combining the different velocity dispersion datasets. The
scatter, $\Delta_{\sigma}$, of the elliptical galaxies is 0.054
compared to 0.073 for the S0 galaxies. This difference may be due to
the fact that S0 galaxies tend to have lower velocity dispersions
which carry with them larger observational errors. We conclude that
there is marginal evidence of a difference between FP relations of E
and S0 galaxies. We have investigated the effects of this difference
on our flow results and find that, because most clusters have similar
proportions of E and S0 galaxies, none of our results are sensitive to
this offset.

\subsection{Stellar population effects and the FP-Mg$_2$ relation} 
\label{sec:mg} 

The presence of a small percentage of stars with intermediate ages
would increase the scatter of the FP relation, and cause systematic
biases in the derived distances. Since the distance-independent Mg$_2$
index is also sensitive to stellar populations, the use of Mg$_2$ as
an additional parameter in the FP relation had been advocated
(G\'uzman \& Lucey 1993). There are, however, several practical
reasons why we prefer not to use Mg$_2$ in our distance indicator. We
shall see that for inverse fits, including Mg$_2$ decreases the
scatter in $\log \sigma$ but actually {\em increases\/} the random
distance error, $\Delta\sb{inv}$ due to the change in slope $\alpha$.
Furthermore, by including Mg$_2$, the total system matching distance
errors increase (by about a third) due to the addition of the Mg$_2$
system matching errors and due to the velocity dispersions system
matching errors, again because of the change in slope $\alpha$.
Finally, only 315 of the 352 galaxies in our sample have Mg$_2$
measurements, and we are required to drop the cluster DC2345-28 which
has only one Mg$_2$ measurement amongst the spectroscopic systems
considered in this paper.

Nevertheless, in order to investigate the effects of stellar 
populations on our results, we have proceeded with fits to an 
FP-Mg$_2$ relation of the form 
\begin{equation} 
\log \sigma = \frac{1}{\alpha} \log R\sb{e} - \frac{\beta}{\alpha} 
\langle\mu\rangle\sb{e} - \frac{\zeta}{\alpha}{\rm Mg}_2 - \frac{1}{\alpha} 
\gamma\sb{cl}\,. 
\label{eq:fpmg} 
\end{equation} 
We find the results $\alpha = 1.795\pm 0.083$, $\beta = 0.331\pm
0.016$ and $\zeta = -2.38\pm0.22$ with $\Delta_{\sigma} = 0.054$ and
$\Delta\sb{inv} = 0.22$. This highly significant detection of the
Mg$_2$ correlation arises as a result of the inverse nature of the
fit. The $\log \sigma$--Mg$_2$ relation has a scatter of only 0.086 in
$\log \sigma$, which is only slightly larger than the scatter in the
inverse FP. The inverse fit thus gives considerable weight to Mg$_2$
as a predictor of $\log \sigma$, in contrast to the forward FP. The
distances of individual clusters typically change by $\la 1\sigma$,
except for Coma which shows an offset at the $2.2\sigma$ level.
Because Coma was our nominal calibrator, all distances require
rescaling by 5\% with the result that Coma has a peculiar velocity of
$+381$ \kms\ and the remaining clusters follow the Hubble flow in the
mean. The best fitting bulk flow of the full cluster sample is in fact
somewhat smaller than that of the FP solution but well within the
$1\sigma$ errors. The motion of the PP ridge towards the LG increases
slightly to $141$ \kms, which again differs by $< 1\sigma$ from the FP
solution.

We conclude that the residuals of individual galaxies from the inverse
FP relation are weakly correlated both with galaxy morphology and with
Mg$_2$ index. However, when averaged over all galaxies in a given
cluster, there is a negligible net effect of such correlations on
derived cluster distances, and hence the flow results are robust
against such effects.

\section{Comparison with other work} 
\label{sec:comp} 
\subsection{Bulk flow and the motion of Perseus--Pisces} 

For the full 16 cluster sample, we find a bulk motion which is
consistent with these clusters being at rest in the CMB frame. In
order to compare our bulk motion, $\bv{V}_1$, with covariance matrix
${\bf C}_1$ with an independent sample with bulk motion $\bv{V}_2$ and
covariance matrix ${\bf C}_2$, we calculate
\begin{equation} 
\chi_{\rm bulk}^2 = (\bv{V}_1 - \bv{V}_2)^{\rm T} \left({\bf C}_1 + {\bf 
C}_2\right)^{-1} (\bv{V}_1 - \bv{V}_2) 
\end{equation} 
and compare the result to a $\chi_{\rm bulk}^2$ distribution with 3 
degrees of freedom (following Hudson \& Ebeling 1997, hereafter HE)% 
\footnote{Note that when comparing two peculiar velocity samples with
  different sky coverage and effective depth, we do not expect the
  measured bulk flows, $\bv{V}_1$ and $\bv{V}_2$, to be identical even
  in the limit of no measurement errors due to the different window
  functions (Watkins \& Feldman 1995). Therefore, the confidence with
  which we conclude that two samples are inconsistent will in general
  be slightly overestimated.}.  We find that the bulk motion of our
sample is consistent with the $360\pm40 $ \kms\ motion found by
Courteau et al.\ (1993). However, the bulk motion found in this paper
is inconsistent at the 98\% confidence level with the result of LP,
who used the photometry of brightest cluster galaxies (BCGs) as a
distance indicator. Our result is also inconsistent at the 94\%
confidence level with bulk motion obtained by HE, who applied a
correction for the X-ray luminosity of the host cluster to the BCG
distance indicator of LP.

We find a negligible mean radial motion for the PP ridge ($-60\pm220$
\kms), which is, at face value, marginally inconsistent with the
result of Willick (1990) who found a mean peculiar velocity of
$-441\pm49$ \kms\ (random error) from a TF study of PP field spirals
in the redshift range $3800 < cz \leq 6000$ \kms. Note, however, that
the systematic calibration error on this result is $\sim 100$ \kms\ 
(Willick 1991).  Furthermore, the two samples probe different parts of
the PP supercluster: the PP ridge clusters studied here range from $0
\hr \la$ RA $\la 3\hr$ whereas the volume probed by Willick covers a
wider range of redshifts and extends farther to the west, from $22 \hr
\la$ RA $\la 3 \hr$. Willick (1991) noted that the western half of his
sample had a greater infall toward the LG.  It is not clear then that
the two results are in significant disagreement.

Our result is also marginally inconsistent with the result of HM who
found a PP radial motion of --400 \kms\ from a TF cluster survey.
However, it should be noted here that the HM and Willick (1990, 1991)
results are not independently calibrated. Both studies use Aaronson et
al. (1986) linewidths for the calibrating cluster sample, and
linewidths from the work of Giovanelli, Haynes and collaborators (e.g.
Giovanelli \& Haynes 1985; Giovanelli et al. 1986) for the PP sample.
HM also adopt the prescription of Willick (1991) to transform between
the two systems. The transformation is based on a sample of 58
galaxies in common between the two linewidth sources.

\subsection{Individual cluster distances and peculiar velocities} 

In order to understand the above-mentioned conflicts between our
results and those of Han \& Mould, Lauer \& Postman and Hudson \&
Ebeling, we now turn to a comparison of distance and peculiar
velocities of individual clusters.

\begin{figure*}\epsfxsize=180mm 
\epsfbox{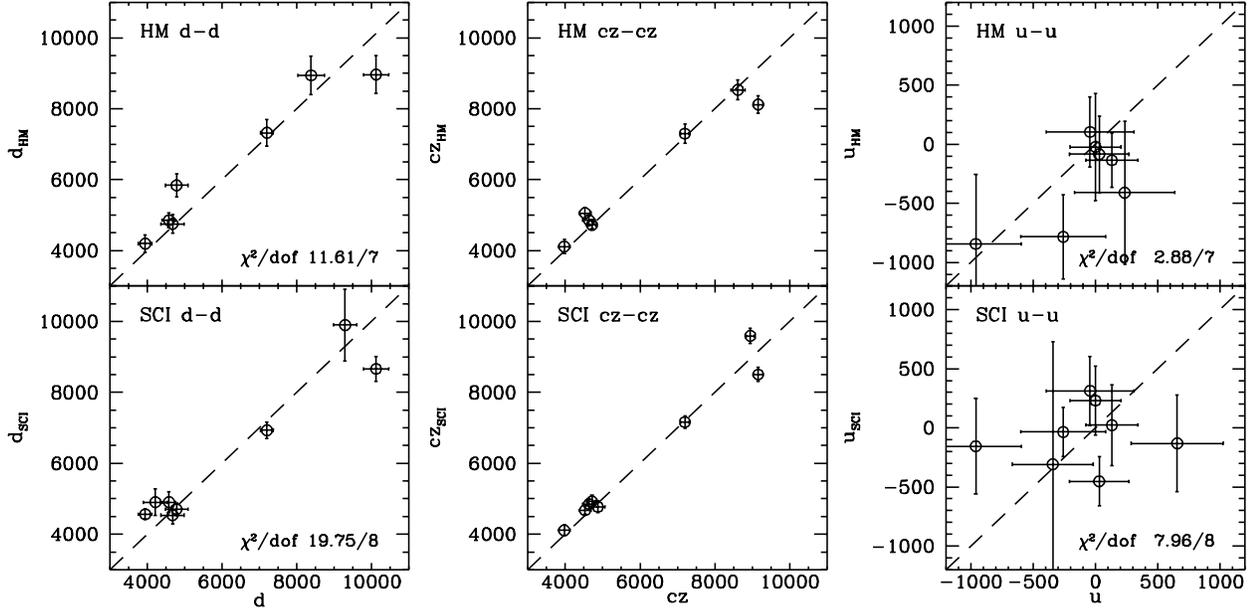} 
\caption{Comparison between our cluster distances and peculiar 
  velocities and literature values for the same clusters. Left-hand
  panels show the distance comparison. The dashed line has slope
  unity. The $\chi^2$ of the comparison and the number of degrees of
  freedom bottom right-hand corner. Middle panels show the comparison
  of cluster mean redshifts. Right-hand panels show comparison of
  cluster peculiar velocities. In all panels, the results of this
  paper are plotted on the horizontal axis. The upper panels show the
  results of HM, as rederived by Willick et al.\ (1997a); and lower
  panels show the results of Giovanelli et al.\ (1997a,b).}
\label{fig:comparetf} 
\end{figure*}

We have compared our cluster distances and peculiar velocities with
the inverse TF cluster distances and peculiar velocities of HM, as
rederived by Willick et al.\ (1997a) for the 7 clusters in common to
the two samples. The results are shown in the upper panels of
\figref{comparetf} and peculiar velocities are tabulated in
\tabref{compare}.

\begin{figure*}\epsfxsize=180mm 
\epsfbox{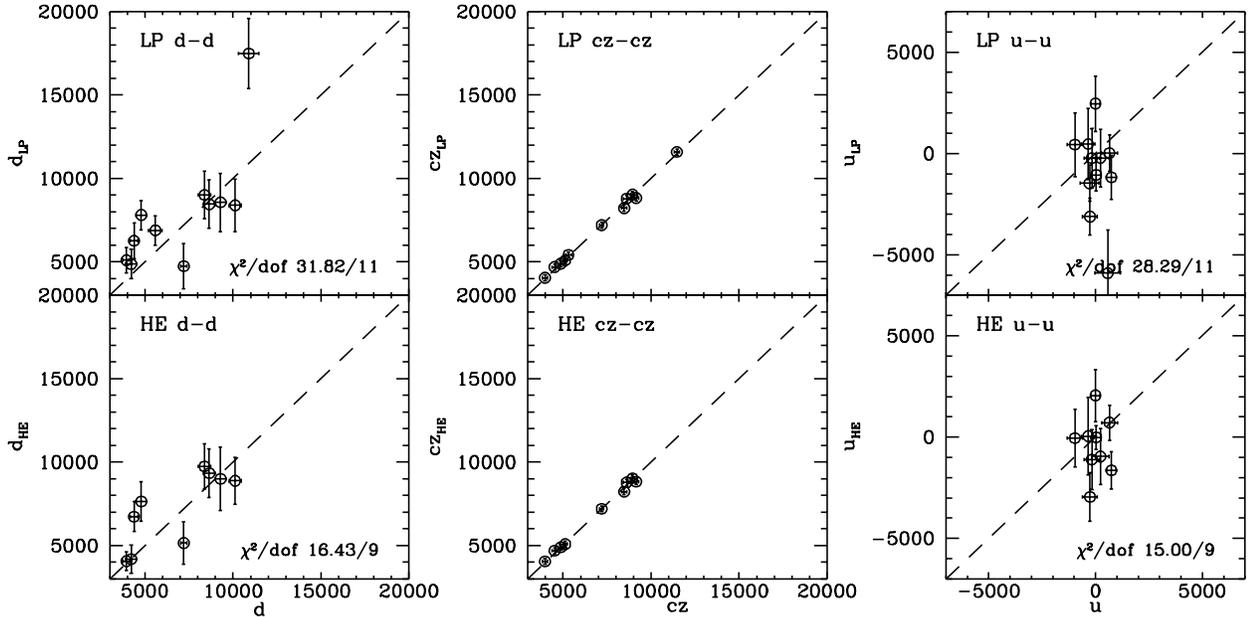} 
\caption{ 
  As in \figref{comparetf}, but note the expanded scale. Upper panels
  show the BCG results of Lauer \& Postman (1994); lower panels show
  the X-ray corrected BCG distances of Hudson \& Ebeling (1997).}
\label{fig:comparex} 
\end{figure*} 

\begin{table*}
\caption{Peculiar velocity comparisons. Our measured cluster 
  peculiar %H
  velocities
  are compared with FP, TF and BCG results from the literature, for
  clusters in common.  Literature sources are JFK96, HM 
  data rederived by Willick et al.\ (1997a), %H 
  Giovanelli et al.\ (1997b) (SCI), LP and HE.  
  To the distance errors we have added in quadrature the error in the %H
  mean redshift, using the number of galaxies in the cluster and the  %H
  cluster velocity dispersions derived in our work.                   %H
  For the BCG comparisons we adopt the mean redshift error of 184     %H
  \kms\ quoted by Postman \& Lauer (1995).                            %H
  Note that our results are very similar to those of JFK96, as expected since
  JFK is the principal data source for 6 clusters in this work.} 
\label{tab:compare}
\begin{tabular}{lr@{$\pm$}rr@{$\pm$}rr@{$\pm$}rr@{$\pm$}rr@{$\pm$}rr@{$\pm$}rl}
&
\multicolumn{4}{c}{Fundamental Plane} &
\multicolumn{4}{c}{Tully--Fisher} &
\multicolumn{4}{c}{Brightest Cluster Galaxy} &
\\
Cluster &
\multicolumn{2}{c}{$u_{\rm PP}$} &
\multicolumn{2}{c}{$u_{\rm JFK}$} &
\multicolumn{2}{c}{$u_{\rm HM}$} &
\multicolumn{2}{c}{$u_{\rm SCI}$} &
\multicolumn{2}{c}{$u_{\rm LP}$} &
\multicolumn{2}{c}{$u_{\rm HE}$} &
\multicolumn{1}{c}{Comments} \\
&
\multicolumn{2}{c}{\kms} &
\multicolumn{2}{c}{\kms} &
\multicolumn{2}{c}{\kms} &
\multicolumn{2}{c}{\kms} &
\multicolumn{2}{c}{\kms} &
\multicolumn{2}{c}{\kms} &\\
Pisces     & $  131$ & $ 208$        & \multicolumn{2}{c}{---} &$ -136$ & $ 229$        & $   23$ & $ 343$        & \multicolumn{2}{c}{---} & \multicolumn{2}{c}{---} &          \\
HMS0122    & $  -44$ & $ 352$        & \multicolumn{2}{c}{---} &$  104$ & $ 298$        & $  313$ & $ 291$        & \multicolumn{2}{c}{---} & \multicolumn{2}{c}{---} &          \\
A262       & $ -259$ & $ 340$        & \multicolumn{2}{c}{---} &$ -783$ & $ 355$        & $  -34$ & $ 206$        & $-3119$ & $ 900$ & $-2953$ & $1208$        &  see text \\
A347       & $ -277$ & $ 430$        & \multicolumn{2}{c}{---} &\multicolumn{2}{c}{---} & \multicolumn{2}{c}{---} & $-1467$ & $ 896$        & \multicolumn{2}{c}{---} &          \\
A2199      & $ -342$ & $ 325$        & \multicolumn{2}{c}{---} &\multicolumn{2}{c}{---} & $ -308$ & $1035$        & $  464$ & $1753$        & $   40$ & $1906$        &          \\
A2634      & $ -960$ & $ 364$        & \multicolumn{2}{c}{---} &$ -844$ & $ 590$        & $ -156$ & $ 405$        & $  437$ & $1577$        & $  -48$ & $1428$        &  see text \\
Coma       & $    0$ & $ 204$        & $    0 $ & $ 148 $      &$  -23$ & $ 454$        & $  231$ & $ 290$        & $ 2459$ & $1372$        & $ 2051$ & $1282$        &          \\
A194       & $  743$ & $ 230$        & $  532 $ & $ 188 $      &\multicolumn{2}{c}{---} & \multicolumn{2}{c}{---} & $-1178$ & $1088$        & $-1643$ & $ 914$        &          \\
A539       & $  234$ & $ 403$        & $  213 $ & $ 285 $      &$ -410$ & $ 604$        & \multicolumn{2}{c}{---} & $ -225$ & $1427$        & $ -956$ & $1376$        &          \\
A3381      & $  578$ & $ 593$        & $  667 $ & $ 698 $      &\multicolumn{2}{c}{---} & \multicolumn{2}{c}{---} & $-5905$ & $2120$        & \multicolumn{2}{c}{---} &          \\
A3574      & $  655$ & $ 369$        & $  556 $ & $ 388 $      &\multicolumn{2}{c}{---} & $ -131$ & $ 408$        & $   20$ & $ 900$        & $  711$ & $ 862$        &          \\
DC2345-28  & $ -174$ & $ 362$        & $ -291 $ & $ 265 $      &\multicolumn{2}{c}{---} & \multicolumn{2}{c}{---} & $ -233$ & $1467$        & $-1108$ & $1481$        &          \\
Hydra      & $   30$ & $ 239$        & $   87 $ & $ 261 $      &$  -85$ & $ 322$        & $ -453$ & $ 208$        & $-1063$ & $ 791$        & $  -17$ & $ 584$        &          \\
\\
\end{tabular}
\end{table*}

The two sets of cluster {\em distances\/} are marginally inconsistent
at the 90\% confidence level. However, we find that there is good
agreement between cluster {\em peculiar velocities\/}. Note that
elliptical and spiral galaxies in the same nominal cluster do not
necessarily sample the same kinematic object. In particular, spiral
cluster samples typically extend to larger radii, and may be more
prone to contamination from the field, or from infalling groups.

For example, for the most discrepant cluster, A262, we find $cz = 4528
$ \kms, $d=4787\pm300$ \kms, and hence $u=-259$ \kms\ whereas Willick
et al.\ (1997a) find $cz = 5057$ \kms, $d=5840\pm322$ \kms\ and hence
$u=-783$ \kms\ Furthermore, the nominal cluster centres differ by
1\degr. It is not clear, therefore, whether our A262 cluster is the
same object as that of HM. A similar situation holds for the next most
discrepant cluster, A2634, in which the treatment of the companion
cluster A2666 affects the comparisons. Scodeggio et al. (1995) present
a full discussion of the contamination problems for this cluster. If
we remove either one of these two clusters from both samples, the
distances are compatible.

In contrast to the TF results of Willick (1990,1991), Courteau et al.\ 
(1993) and HM, the recent I-band Tully-Fisher cluster (Giovanelli et
al.\ 1997a, b, hereafter SCI) and field (Giovanelli et al.\ 1996) data
suggest no net motion for PP. The lower panels of \figref{comparetf}
compare our distances with the cluster distances of Giovanelli et al.\ 
(1997a,b). We have computed the mean CMB redshifts of cluster members
from Giovanelli et al.\ (1997a) using galaxies denoted by ``c'' in
column 7 of their Table 2, corresponding to their ``in'' samples) and
peculiar velocities determined from their incompleteness-corrected TF
magnitude offsets from column 5 of Table 3 of Giovanelli et al.\ 
(1997b). Note that their NGC 383 Group corresponds to Pisces and their
NGC 507 group corresponds to HMS0122. While the two samples of cluster
distances disagree at the 99\% confidence level, this disagreement is
mainly due to A2634 for which the mean redshifts differ by 650 \kms.
If this cluster is removed from the comparison, the distances are
consistent.

The data used to determine the peculiar velocities of A2199 and A2634
are the same as that used by LGSC. Whereas we derive values of
$-342\pm325$ \kms\ and $-960\pm364$ \kms\ respectively, LGSC derived
$-160\pm380$ \kms\ and $-670\pm490$ \kms. These relatively small
differences arise from the application of the FP distance indicator.
In this paper we have used the `inverse' relation whereas LGSC used
the `forward' relation with an allowance for the different selection
functions in the clusters.

Scodeggio, Giovanelli \& Haynes (1996) have recently reported new FP
results for Coma and A2634. Using Coma as the calibration cluster,
they derive a distance for A2634 of $9099\pm266$ \kms. While their
photometric parameters are from a new set of I-band data, they used
velocity dispersion data from several sources. For the Coma cluster
these are mostly taken from the literature, while for A2634 they use
mainly new velocity dispersion measurements from the Hale 5m
telescope. They have only a small overlap between their measurements
and the literature values. Their quoted error in the A2634 distance
does not include the uncertainty in linking their velocity dispersion
values onto the `literature' system. In PPI we showed that different
spectroscopic datasets typically differ at the level of 0.01 dex.
Using the new measurements from PPI and LGSC, we find that the
Scodeggio et al. velocity dispersions are $\sim0.02$ dex smaller than
our standard system. While a more extensive analysis is required to
place the Scodeggio et al. data accurately onto a `standard' system,
such a systematic offset would translate into increasing their derived
distance for A2634 by $\sim$500 \kms. After allowing for this
correction the Scodeggio et al. distance for A2634 is not in conflict
with the value reported here, i.e. $\sim9600\pm350$ \kms\ versus
$10118\pm339$ \kms.

Our non-detection of a bulk motion is in disagreement with the result
of LP, who studied BCGs in 119 Abell/ACO clusters within $cz < 15000 $
\kms. They used a distance indicator based on the photometry of BCGs
and concluded that their sample had a bulk motion with respect to the
CMB frame of $689\pm178$ \kms\ towards \lb{343}{52}. The 16 cluster
sample studied here is much smaller than theirs, but the distance
error per cluster is typically much smaller in this work. The results
of comparing individual cluster distances for the 11 clusters in
common are shown in the upper panels of \figref{comparex}. The two
sets of cluster distances (peculiar velocities) disagree at the 99.9\%
(99.7\%) confidence level, which indicates that the errors on one or
both of the data sets are underestimated. Note that, in contrast to
the comparison between spiral and elliptical cluster samples, the mean
redshifts and distances of our samples of cluster ellipticals, which
typically include the BCG, are expected to be very similar to those of
the BCGs alone (see the middle panels of \figref{comparex}).

The lower panels of \figref{comparex} show the comparison between the
clusters of this paper and BCG distances rederived by HE for the 9
clusters in common. The BCG distance indicator of HE is similar to
that of LP but includes a correction to the BCG magnitude for the
X-ray luminosity of the host cluster, because more X-ray luminous
clusters tend to have brighter BCGs. The agreement between the our
cluster distances with those of HE is better than with those of LP,
but is still marginal: the two samples disagree at the 94\% confidence
level. Part of the improved agreement results from the absence of
A3381 from the X-ray sample, and part of the improvement is due to the
X-ray correction itself: for the 9 clusters in common, the reduced
$\chi^2$ is 2.32 for the LP distance indicator (still incompatible at
the 98.7\% confidence level) and 1.83 for the X-ray corrected BCG
distance indicator. Nevertheless, there remain outlying clusters in
the BCG samples: for both A262 and Coma (A1656) both the LP BCG
distance indicator and the X-ray corrected BCG distance indicator of
HE indicate peculiar velocities in excess of 2000 \kms\ (see
\tabref{compare}).

\section{Summary} 
\label{sec:summary} 

We have measured the mean peculiar motions of 103 early-type galaxies
in 7 clusters in the PP region, and a further 249 such galaxies in 9
calibrating clusters from the literature, using the inverse
Fundamental Plane relation. This relation is found to have a distance
error of 20\% per galaxy. Our principal results are as follows:

\begin{enumerate} 
\item{Of 6 clusters in the PP ridge, none shows a significant motion
    with respect to the CMB frame. For the PP background cluster J8,
    there is marginal evidence for `backside infall' into the PP
    supercluster.}
\item{ The PP supercluster has an insignificant net radial motion
    ($-60\pm220$ \kms) \wrt\ the CMB frame.}
\item{An all-sky sample comprised of 16 clusters (with median depth
    $cz \sim 5500$ \kms) exhibits a bulk motion of $420\pm280$ \kms\ 
    towards \lb{262}{-25}. }
\item{Comparison of observed cluster velocities with predictions from
    the IRAS 1.2Jy redshift survey yields $\beta_I \equiv
    \Omega^{0.6}/b_I = 0.95\pm0.48$, consistent with previous
    results.}
\end{enumerate} 

Our error analysis fully accounts for the uncertainties in the mean
Hubble flow as well as the errors due to the merging of different
spectroscopic systems.

The bulk motion of the 16 cluster sample is consistent with the sample
being at rest in the CMB frame, but is also consistent with the $\sim
350$ \kms\ motion found by Courteau et al. (1993). It is inconsistent
with the $\sim 700$ \kms\ bulk motion found by LP.

Our mean PP radial motion result is in apparent conflict with the TF
results of Willick (1990, 1991) and HM which had found a $\sim -400$
\kms\ peculiar velocity of the PP ridge, but our results are in better
agreement with the recent TF data of Giovanelli et al.\ 
(1996,1997a,b).  Comparison of elliptical and spiral samples, is far
from straightforward, however, due to the complex nature of the
peculiar velocity field and the different regions probed by different
surveys.

The disagreement between our cluster distances and those of LP for the
11 clusters in common is statistically significant at the $\ga 99.7$\%
confidence ($\sim 3\sigma$) level indicating that the errors of one or
both of these data sets are underestimated. When the X-ray corrected
BCG distances of HE are used, the disagreement is reduced to the $\sim
94$\% ($\sim 2\sigma$) confidence level.

The cluster peculiar velocity comparison is potentially the cleanest
method of determining $\beta$ on linear scales because it is
relatively insusceptible to Malmquist and smoothing biases. However, a
larger sample of clusters is required to reduce the random errors.

The results from several large-scale peculiar velocity surveys are
expected shortly. Of particular interest are the elliptical sample of
the EFAR collaboration (Wegner et al.\ 1996), the cluster
Tully--Fisher survey of Giovanelli, Haynes and collaborators (Dale et
al.\ 1997) and the Tully-Fisher field survey discussed by Strauss
(1997; see also references therein for other peculiar velocity
surveys).

Finally, with collaborators, three of the present authors are
currently obtaining FP data for ellipticals in an all-sky sample of
approximately 50 clusters within a distance of 12000 \kms. This will
allow a precise measurement of the bulk motion on these scales as well
as a clean determination of $\beta$ in the linear regime.

\section*{Acknowledgments} 

Michael Strauss is thanked for supplying to us the IRAS 1.2Jy density
field. This work has made use of Starlink facilities at Durham. JS and
RJS acknowledge financial support from the PPARC.  MJH acknowledges
financial support from the PPARC; from a CITA National Fellowship; and
from the Sciences and Engineering Research Council of Canada, through
operating grants to F. D. A. Hartwick and C. J. Pritchet.

\end{document}